\documentclass[english,preprint]{aastex}
\usepackage[T1]{fontenc}
\usepackage[latin9]{inputenc}
\usepackage{color}
\usepackage{amsbsy}
\usepackage{amstext}
\usepackage{amssymb}
\usepackage{esint}
\usepackage{babel}
\begin{document}

\title{Dicke's Superradiance in Astrophysics. I -- The 21 cm Line}

\author{Fereshteh Rajabi$^{1}$ and Martin Houde$^{1,2}$}

\affil{$^{1}$Department of Physics and Astronomy, The University of Western
Ontario, London, ON, N6A 3K7, Canada}

\affil{$^{2}$Division of Physics, Mathematics and Astronomy, California
Institute of Technology, Pasadena, CA 91125, USA }
\begin{abstract}
We have applied the concept of superradiance introduced by Dicke in
1954 to astrophysics by extending the corresponding analysis to the
magnetic dipole interaction characterizing the atomic hydrogen 21
cm line. Although it is unlikely that superradiance could take place
in thermally relaxed regions and that the lack of observational evidence
of masers for this transition reduces the probability of detecting
superradiance, in situations where the conditions necessary for superradiance
are met (i.e., close atomic spacing, high\textbf{ }velocity coherence,
population inversion, and long dephasing time-scales compared to those
related to coherent behavior), our results suggest that relatively
low levels of population inversion over short astronomical length-scales
(e.g., as compared to those required for maser amplification) can
lead to the cooperative behavior required for superradiance in the
ISM.\textbf{ }Given the results of our analysis, we expect the observational
properties of 21-cm superradiance to be characterized by the emission
of high intensity, spatially compact, burst-like features potentially
taking place over short\textbf{ }periods ranging from minutes to days. 
\end{abstract}


\section{Introduction\label{sec:Introduction}}

It is generally assumed that in much of the interstellar medium (ISM)
emission emanating from atomic and molecular transitions within a
radiating gas happen independently for each atom or molecule. From
intensity measurements of such spectral lines, important parameters
(e.g., density and temperature) can be determined and the physical
conditions in a given\textbf{ }environment thus characterized\textbf{
}\citep{Townes1955, Emerson1996, Goldsmith1999, Irwin2007}. For example,
in cases where the spectral lines are optically thin the intensity
will be found to scale linearly with the number of atoms or molecules
responsible for the detected radiation. The soundness of this approach
rests mostly on the assumption that spontaneous emission from different
atoms or molecules happens independently. 

As was pointed out by R. H. Dicke in a seminal paper several decades
ago \citep{Dicke1954}, the assumption of independent spontaneous
emission for the components of a gas does not apply in all conditions.
As will be discussed in this paper, and following Dicke's original
analysis, closely packed atoms can interact with their common electromagnetic
field and radiate coherently. That is, the spontaneous emission of
atoms or molecules in such a gas will not be independent, but rather
take place in a cooperative manner. In the ideal case, this phenomenon
will lead to a much more intense and focussed radiation (proportional
to the square of the number of atoms), which Dicke called superradiance.
Since Dicke's original proposal, the field of superradiance research
has flourished and an abundant literature has developed within the
physics community. The first experimental detection of superradiance
in the laboratory was achieved by \citet{Skribanowitz1973}, while
several other independent verifications \citep{Gross1976, Gibbs1977, Carlson1980, Moi1983, Greiner2000, Xia2012}
have since been realized under a large domain of conditions and experimental
setups (see Chap. 2 of \citealt{Benedict1996, MacGillivray1976, Andreev1980, Gross1982}
for reviews). 

While the reality of the superradiance phenomenon has long been clearly
established in the laboratory, to the best of our knowledge, it has
yet to be investigated within an astrophysical context. It appears
to us important to do so since some of the requirements and conditions
needed for the realization of a superradiant system are known to be
satisfied in some regions of the ISM. More precisely, superradiance
can arise in systems where there is a population inversion, and the
effect will be much stronger and more likely to be realized when atoms
or molecules are separated by approximately less than the wavelength
of radiation (see below and Section \ref{sub:Large-N-atom-Sample}). 

The population inversion condition is known to occur in the ISM and
is partly responsible for the ubiquitous presence of masers (see \citealt{Fish2007,Watson2009,Vlemmings2012,Sarma2012}
for recent reviews). But it is also important to realize that, although
it is a necessary condition, population inversion is not by itself
sufficient to ensure superradiance. It is also required that there
exists sufficient velocity coherence between the atoms partaking in
the effect, and that any other dephasing takes place on time-scales
longer than those characterizing superradiance. When all these conditions
are met a coherent behavior can be established between the atoms and
superradiance can ensue. We note, however, that, as will be discussed
later on, superradiance is unlikely to could take place in thermally
relaxed regions of the ISM. This is because Doppler broadening resulting
from, say, a Maxwellian velocity distribution would leave too few
atoms with the required velocity coherence to allow superradiance
to develop. Our analysis will therefore imply other types of environments
where thermal equilibrium has not been reached. For example, any region
in the ISM into which a significant amount of energy is being suddenly
released (e.g., shocks or regions where significant radiation flares
occur) will be strongly out of equilibrium, and provide conditions
that are potentially markedly different to those found in a thermal
gas and may meet the requirements for superradiance. Also, although
superradiance can also occur for large interatomic or molecular separations
(i.e., greater than the wavelength of radiation; see Section \ref{sub:Two-Atom-large}),
the aforementioned constraint of small interatomic or molecular separation,
and its implication for the corresponding densities, is likely to
be met for only a limited number of spectral lines, but a few astrophysically
important transitions are suitable candidates. One of these spectral
lines is the 21 cm atomic hydrogen transition. 

Even though a 21 cm maser has yet to be discovered, which would also
imply the realization of a population inversion for this spectral
line, as will be seen through our analysis the length-scales required
for superradiance at 21 cm are very small compared to those that would
be needed for maser amplification in the ISM (\citealt{Storer1968},
and see below). It follows that although the lack of observational
evidence of masers for this transition significantly affects the probability
of detecting superradiance, it does not rule it out. Also, the existence
of higher densities of atomic hydrogen in some parts of the ISM would
increase the potential detectability of superradiance, if the other
necessary conditions for its realization previously listed were also
met. Furthermore, with the recent discoveries of radio bursts at frequencies
close to 1400 MHz \citep{Kida2008, Thornton2013} the investigation
of the properties of a transient phenomenon such as superradiance
is timely. This is why in this first paper on the subject we chose
to introduce the concept of superradiance to the ISM using this spectral
line. 

Whether or not a population inversion can easily be realized for the
energy levels leading to the 21 cm line , it has been considered in
the existing literature \citep{Shklovskii1967,Storer1968,Dykstra2007}
and we know of at least one region (the Orion Veil) where the kinetic
temperature is lower than the 21-cm spin temperature, providing evidence
for a population inversion \citep{Abel2006}. The main pumping process
covered in the literature corresponds to the situation when a H{\footnotesize I}
gas is close to a source of radiation that emits a field with an intensity
$I_{v}\left(\nu\right)$ in the neighborhood of the Lyman $\alpha$
line. A hydrogen atom in the ground hyperfine state ($n=1,\:F=0$)
can absorb a photon and become excited to the $n=2$ level. Later
on, the atom returns to the upper hyperfine state ($n=1,\:F=1$) emitting
a slightly less energetic photon than the initial one absorbed by
the atom. The same can happen for a hydrogen atom initially in the
hyperfine state ($n=1,\:F=1$) that returns to the ground ($n=1,\:F=0$)
state after excitation to the $n=2$ level, emitting a slightly more
energetic photon in the process. The absorption rate of the photons
for both cases depends on the intensity of the radiation $I_{v}\left(\nu\right)$,
but the return (emission) process does not. Therefore, the $F=0$
level will undergo more absorptions followed by a return to the ($n=1,\:F=1$)
level whenever $I_{v}\left(\nu\right)$ harbors more blue than red
photons, and will become accordingly less populated than the $F=1$
level \citep{Wouthuysen1952,Field1958,Shklovskii1967,Storer1968}.
Although \cite{Storer1968} concluded that it is unlikely to maintain
a population inversion over an extended region needed for the maser
amplification with this process, they also pointed out that an ``appreciable''
inversion can thus be realized over a region of thickness $\sim6\times10^{-5}$
pc. Given the above inversion scenario, we would expect that environments
located in the periphery or near boundaries of H{\footnotesize II}
regions could provide conditions suitable for the development of superradiance,
for example. The aforementioned evidence for a 21 cm population inversion
in the Orion Veil brings support to this idea. Whatever the case,
the 21 cm line will serve us as a starting point for the development
of the superradiance formalism for the ISM (in the present case for
a magnetic dipolar transition), which will then be refined in the
future and also applied to other (electric dipolar) spectral lines
(e.g., the OH 1612-MHz, CH$_{3}$OH 6.7-GHz, and H$_{2}$O 22-GHz
maser transitions) where observational evidence for superradiance
can be found in the literature \citep{Rajabi2016a, Rajabi2016b}.

It should also be pointed out that superradiance is a fundamentally
different phenomenon from the maser action, even though the two may
seem similar at first glance. An astronomical maser is a collective
but not coherent phenomenon. More precisely, for a maser a group of
atoms, initially in their excited states, emit through the stimulated
emission process but cannot be considered as a single quantum system.
That is, it is possible to describe maser action through successive
events where an excited atom is stimulated by the incident radiation
and emits a photon, with the same stimulation/emission processes subsequently
repeated for different atoms in the masing sample. On the other hand
for superradiance, coherence emphasizes the fact that the group of
atoms interacting with the radiation field behaves like a single quantum
system \citep{Nussenzveig1973}. That is, the superradiance emission
process cannot be broken down into successive events as is the case
of maser radiation. Finally, superradiance is a transient effect in
which a strong directional pulse is radiated over a relatively short
time-scale, while maser action operates more in a steady state regime
as long as population inversion is maintained.

The material covered in this paper goes as follows, we start with
a general discussion of the concept of superradiance for the so-called
small- and large-samples, as originally discussed by \citet{Dicke1954, Dicke1964},
in Section \ref{sec:Superradiance}. In Section \ref{sec:The-two-level-H-sample},
we examine the possibility of building cooperative behavior in a H{\footnotesize I}
sample based on a comparative analysis of time-scales for the 21 cm
line in a H{\footnotesize I} gas, as well as present corresponding
numerical results. A discussion and short conclusion follow in Sections
\ref{sec:Discussion} and \ref{sec:Conclusion}, respectively, while
the superradiance formalism and detailed derivations for the material
discussed in the main sections of the paper will be found in appendices
at the end.

\section{Superradiance \label{sec:Superradiance}}

\subsection{Dicke's Small-sample Model\label{sub:Dicke's-Small-sample-Model}}

\begin{figure}[tb]
\epsscale{0.3}\plotone{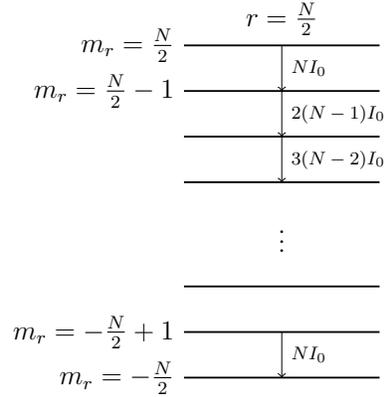}

\caption{\label{fig:cascade}Dicke states with $r=N/2$ for a system of $N$
two-level (spin-1/2) particles. Spontaneous radiation intensities
are indicated on the right.}
\end{figure}

Dicke originally proposed in 1954 a model where an ensemble of $N$
initially inverted two-level atoms interacting with their common radiation
field is considered as a single quantum mechanical system \citep{Dicke1954}.
In his model, a two-level atom is modeled as a spin-1/2 particle in
a magnetic field where the spin up configuration corresponds to the
excited state $|e\rangle$ and the spin down to the ground state $|g\rangle$.
Just as an ensemble of $N$ spin-1/2 particles can be described using
two quantum numbers $s$ and $m_{s}$, the eigenstates of the combined
$N$ two-level atoms in Dicke's model can also be labelled with two
quantum numbers $r$ and $m_{r}$ such that $0\leq r\leq N/2$ and
$m_{r}=-r,-r+1,\ldots,r-1,r$, where

\noindent 
\begin{equation}
m_{r}=\frac{N_{e}-N_{g}}{2},\label{eq:mr}
\end{equation}

\noindent with $N_{e}$ and $N_{g}$ the number of particles in the
excited and ground states, respectively. From the complete set of
eigenstates characterizing this quantum mechanical system, those symmetrical
under the permutation of any pair of atoms are particularly important
and are called Dicke states. The initial state $|e,e,...,e\rangle$
of $N$ fully inverted spin-1/2 particles corresponding to $N$ fully
inverted two-level atoms is one such Dicke state, and is identified
by $r=N/2$ and $m_{r}=N/2$. When an atom in the ensemble decays
to its ground state by emitting a photon, the quantum number $m_{r}$
is decreased by one while $r$ remains unchanged and the system moves
to another symmetric state. Dicke showed that the radiation intensity
from such an ensemble cascading from the initial ($r=N/2,\:m_{r}=N/2$)
state down through an arbitrary state ($r,\:m_{r}$) is 

\begin{equation}
I=I_{0}\left(r+m_{r}\right)\left(r-m_{r}+1\right)\label{eq:I_Dicke}
\end{equation}

\noindent if the volume containing the ensemble of $N$ two-level
atoms is much smaller than $\lambda^{3}$, the cube of the wavelength
of the radiation interacting with the atoms. In Equation (\ref{eq:I_Dicke}),
$I_{0}$ is the radiation intensity due to spontaneous emission from
a single two-level atom. This particular type of system and density
condition defines a \emph{small-sample}. This cascading process is
depicted in Figure \ref{fig:cascade}.

Furthermore, Dicke pointed out that in the ($r=N/2$, $m_{r}=0$)
state, where the half of the atoms are in the ground state and the
other half in the excited state, the radiation intensity of the system
is maximum at

\begin{eqnarray}
I & = & I_{0}\left(\frac{N}{2}\right)\left(\frac{N}{2}+1\right)\label{eq:I_Dicke_max}\\
 & \propto & N^{2}I_{0},\label{eq:I_Dicke_N2}
\end{eqnarray}

\noindent implying a significantly enhanced radiation beam, a phenomenon
he named \emph{superradiance}. This can be understood by the fact
that when the distance between neighboring atoms is much smaller than
the wavelength of radiation, the photon emitted by one atom is seen
to be in phase by neighboring atoms and can bring about the emission
of a new photon of the same mode and in the same direction as the
initial photon. This process can continue through the whole ensemble
resulting in an intense superradiant radiation pulse proportional
to $N^{2}$ (see Equation {[}\ref{eq:I_Dicke_N2}{]}). In contrast
to superradiance observed in a perfectly coherent system, in a non-coherent
system all atoms act independently with a radiation intensity scaling
linearly with $N$. This possibility of coherent interactions is in
contrast with the common assumption that in the ISM atoms, for example,
mainly interact independently with the radiation field, such that
the intensity of the radiation is a linear function of the atomic
density.

In order to conduct a more careful investigation of the possibility
of coherent interactions, especially superradiance in a H{\footnotesize I}
gas, we will need to adapt Dicke's original theory to the corresponding
astrophysical conditions. We therefore first need to carefully understand
all the assumptions that lead to a symmetrical ensemble and superradiance
in the original model of \citet{Dicke1954}. The main assumptions
can be listed as follows:
\begin{itemize}
\item A small-sample of neutral atoms is confined to a volume $\mathcal{V}\ll\lambda^{3}$
with the walls of the volume transparent to the radiation field.
\item The $N$ two-level atoms in the sample are separated by a distance
much less than $\lambda$ but distant enough not to worry about any
overlap between the wave functions of neighboring atoms, which would
require that the wave functions be symmetrized.
\item The ensemble of $N$ initially inverted hydrogen atoms possesses a
permutation symmetry under the exchange of any pair of atoms in the
sample. This is a restricting condition that could prove difficult
to satisfy in general.
\item The transition between atomic levels takes place between non-degenerate
levels, collisions between atoms do not affect their internal states
and collisional broadening is neglected as a result of the small size
of the sample \citep{Dicke1953}.
\item Although it is mentioned in \citet{Dicke1954} that the main results
of his study are independent of the type of coupling between atoms
and the field, the interaction of the atoms with the radiation field
in Dicke's model is assumed to be electric dipolar. 
\item Finally, the radiation field is assumed to be uniform through the
small-sample, the electric dipoles associated to the atoms are parallel,
and propagation effects neglected. 
\end{itemize}
Comparing a corresponding small-sample of $N$ neutral hydrogen atoms
interacting with 21 cm line in the ISM with a Dicke sample, we can
see that some of the assumptions made in the Dicke formalism hold
and some do not. For example, the transitions between the hyperfine
states of a hydrogen atom take place between non-degenerate levels
since the external magnetic field in the ISM lifts the upper level
degeneracy (see Section \ref{sec:The-two-level-H-sample}). Also,
a small-sample of H{\footnotesize I} atoms found in many regions
in the ISM would readily verify the criterion that $N\gg1$ in a volume
$\mathcal{V}<\lambda^{3}$, and could thus be approximately assumed
to experience the same 21-cm radiation field without consideration
of propagation effects. On the other hand, unlike in Dicke's sample
collisional and Doppler broadening effects should, in the most general
case, be considered because, for example, collisions between hydrogen
atoms affect the internal hyperfine states in their electronic ground
state through spin de-excitation \citep{Field1958}. Most importantly,
it must also be noted that the type of coupling between hydrogen atoms
and the 21 cm line is magnetic dipolar in nature.

Above all, the permutation symmetry of atoms, which is a key assumption
in the Dicke model, is difficult to be preserved in an actual situation
because of dipole-dipole interactions between the atoms. Dipole-dipole
interactions have a $r^{\prime-3}$ dependency and these short-range
interactions become important in small-samples where the distance
between atoms $r^{\prime}$ is smaller than $\lambda$ (see Section
\ref{sec:magnetic-dipole-dipole} below). In the Dicke model, the
symmetry breaking effect of dipole-dipole interactions is ignored.
In later studies of superradiance (e.g., \citealt{Gross1982}), it
has been shown that in general dipole-dipole interactions break the
permutation symmetry except in those configurations where all atoms
have identical close-neighbor environments. This symmetry breaking
effect results in weakened correlations and a subsequent deviation
from a perfectly symmetrical superradiance behavior (i.e., the $I\propto N^{2}$
relation in Equation {[}\ref{eq:I_Dicke_N2}{]}). In a sample of $N$
atoms, if $s$ atoms ($s<N$) experience a similar close-neighborhood,
the correlation can build-up among this group of atoms and the intensity
of radiation from the whole sample is expected to be larger than the
intensity of a fully non-coherent system ($I_{\mathrm{nc}}$) but
smaller than that of a perfect superradiance system ($I_{\mathrm{SR}}$). 

In a small-sample of $N$ neutral hydrogen atoms in the ISM it may
thus appear possible to develop coherent behaviors if the permutation
symmetry is conserved among a group of atoms in the sample. This is
arguably a reasonable assumption on average for an ensemble of atoms
within the small volumes discussed here. That is, the different atoms
in the sample are likely to be subjected to the same conditions when
averaged over time and space. Furthermore, we also note that in a
H{\footnotesize I}-sample the magnetic dipole-dipole interactions
are definitely weaker than the electric dipole-dipole interactions
discussed in the literature focusing on symmetry breaking effects.

\subsection{Dicke's Large-sample Model\label{sub:Dicke's-Large-sample-Model}}

In his first paper on superradiance, Dicke also extended his formalism
to a large-sample, where the volume of the sample $\mathcal{V}>\lambda^{3}$
and the interatomic distance $r^{\prime}$ between some atoms can
be greater than $\lambda$. He showed that in a large-sample, coherent
radiation can occur in a particular direction ${\bf k}$ in which
the radiation from different atoms are in phase. When the phase-matching
condition is satisfied in some direction ${\bf k}$, the initial state
of the system can be described by a correlated symmetric state of
type $\left(r,m_{r}\right)$, and the intensity of the radiation in
a solid angle along $\mathbf{k}$ follows 

\begin{eqnarray}
I\left(\mathbf{k}\right) & = & I_{0}\left(\mathbf{k}\right)\left[\left(r+m_{r}\right)\left(r-m_{r}+1\right)\right],\label{eq:large-sample-dicke-intensity}
\end{eqnarray}
similar to Equation (\ref{eq:I_Dicke}) for a small-sample. When a
photon is emitted in the direction ${\bf k}$, the system cascades
to a lower state obeying the selection rules $\Delta r=0,\,\Delta m_{r}=-1$,
and similar to the case of a small-sample, symmetrical states of the
same $r$ are coupled to each other through coherent transitions (see
Section \ref{sub:Two-Atom-large}). On the other hand, when a radiated
photon has a wave vector ${\bf k'\neq{\bf k}}$, the states with different
$r$ (i.e., of different symmetry) can couple and consequently the
coherence is weakened in the system \citep{Dicke1954}. It follows
that in a large-sample consisting of $N$ inverted atoms, the radiation
by one atom is only seen to be in phase by a group of atoms (contrary
to a small-sample where the radiation field is assumed uniform over
the whole sample), and correlation can only be developed among this
group. This naturally results in a radiation intensity that is greater
than that of the corresponding fully non-coherent system but smaller
than the superradiance intensity of a perfectly coherent system consisting
of $N$ atoms. 

Finally, in a large-sample as a result of possibly large interatomic
distances (i.e., $r'>\lambda$) the symmetry breaking effects of the
dipole-dipole interactions are less important, whereas, the propagation
effects that are absent in a small-sample cannot be neglected. The
propagation of radiation over a large distance in a large-sample results
in the re-absorption and re-emission of the photons and consequently
leads to a non-uniform evolution of the atoms in the sample (see Section
\ref{sub:Large-N-atom-Sample}). Beyond these factors, Dicke's analysis
of the large-sample includes similar assumptions as those used for
the small-sample.

\section{The Two-level H{\footnotesize I}-sample \label{sec:The-two-level-H-sample}}

Let us consider an ensemble of neutral hydrogen atoms in the electronic
ground state in some region of the ISM, where it can emit or absorb
photons at the $\lambda=21$ cm wavelength. The hydrogen 21 cm line
is perhaps the most important source of information in radio astronomy
and arises from the transition between two levels of the hydrogen
atom in the 1s ground state. The interaction between the electron
spin and the proton spin in the nucleus of the atom splits the otherwise
degenerate 1s energy level into the two $F=0$ and $F=1$ sub-levels.
The $F=1\leftrightarrow0$ transition in the absence of an external
magnetic field produces the 21 cm line corresponding to a frequency
$\nu=1420.406$ MHz. 

\begin{figure}[tb]
\epsscale{0.5}\plotone{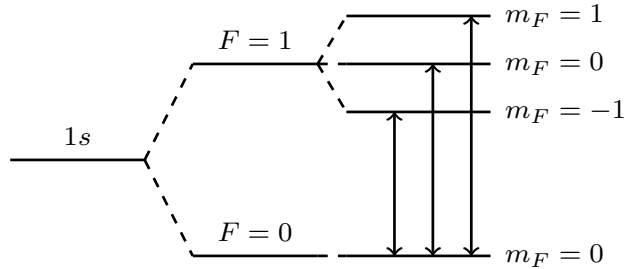}

\caption{\label{fig:leveldiagram}Energy level diagram for the H{\footnotesize I}
21 cm line in the presence of a Zeeman-splitting external magnetic
field. }
\end{figure}

Considering a more realistic case, the magnetic field in a cold neutral
gas is generally on the order of 10 $\mu$G \citep{Crutcher2012},
and the energy level corresponding to $F=1$ splits into three sub-levels
identified by $m_{F}=-1,0,\:\mathrm{and}\:1$. The interaction between
the $F=0$ and $F=1$ levels becomes more complicated as this splitting
provides three possible hyperfine transitions, as shown in Figure
\ref{fig:leveldiagram}. These hyperfine transitions link states of
like parity and obey the general magnetic-dipole selection rules $\Delta F=0,\pm1$
and $\Delta m=0,\pm1$. Based on these rules, all of the three transitions
shown in Figure \ref{fig:leveldiagram} are allowed, however, depending
on the relative orientation (or the polarization) of the magnetic
component of the radiation field to the quantization axis of the atom,
some transitions may be favored. In the more general case, there is
a mixture of all three transitions with each transition exhibiting
particular polarization properties. In order to better understand
the coherent and cooperative evolution of a sample of $N$ hydrogen
atoms coupled to its radiation field, it will be simpler for us to
focus our analysis on only one of these transitions and consider the
atomic system as an ensemble of two-level atoms. Although this model
represents a significant simplification, the two-level atom approximation
is extensively used for, and its results well-verified in, laboratory
experiments involving more complicated atomic or molecular systems
with more complex energy levels \citep{Mandel2010}.

\subsection{Magnetic Dipole-dipole Interaction Between Hydrogen Atoms\label{sec:magnetic-dipole-dipole}}

The theoretical model for the problem will be found in Appendix \ref{sec:Theoretical-Model},
where the Hamiltonian for the two-level H{\footnotesize I}-sample
is developed and the main equations of superradiance derived. To simplify
our discussion we have limited our analysis to the $\left|F=0,\:m=0\right\rangle \longleftrightarrow\left|F=1,\:m=+1\right\rangle $
transition through which a hydrogen atom emits a left circular polarization
(LCP) photon, i.e., with its electric field vector rotating counter-clockwise
as seen by the observer facing the incoming wave. One of the main
components of the Hamiltonian is the magnetic dipole energy term $\hat{V}_{\mathrm{MD}}$
that describes the interaction betweens the atoms composing the sample
(see Equations {[}\ref{eq:V_MD}{]} and {[}\ref{eq:V_MDquantized}{]}).
We now focus on this interaction to get a sense of how the needed
cooperative behavior for superradiance develops between atoms. 

\begin{figure}[h]
\epsscale{0.45}\plotone{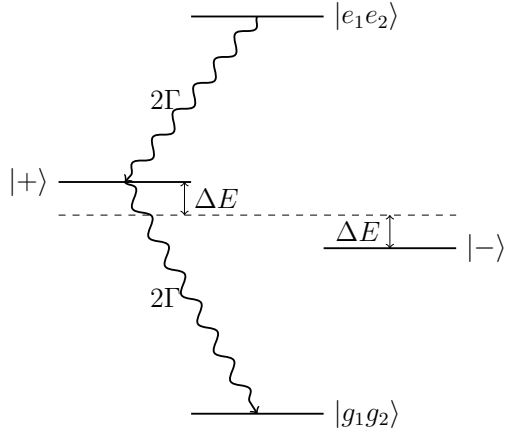}

\caption{\label{fig:two-atoms}The two-hydrogen-atom system. When $kr^{\prime}\ll1$
the upper and lower symmetric states $\left|e_{1}e_{2}\right\rangle $
and $\left|g_{1}g_{2}\right\rangle $, respectively, couple to the
intermediate symmetric state $\left|+\right\rangle $ at the enhanced
transition rate $2\Gamma$, where $\Gamma$ is the transition rate
of a single atom acting independently. In contrast, the antisymmetric
state $\left|-\right\rangle $ cannot couple to the upper and lower
states because of the cooperative behavior between the two atoms.
The energy level shifts $\pm\Delta E$ for the $\left|\pm\right\rangle $
states are also shown. }
\end{figure}

\subsubsection{Hydrogen Atoms Separated by a Small interatomic Distance $\left(r'<\lambda\right)$\label{sub:two-atom-small}}

In order to have a better understanding of how cooperative behavior
is built up in a sample of $N$ atoms, it is helpful to first study
the simpler case of two atoms. We specifically consider a system consisting
of two hydrogen atoms, once again assuming each atom is a two-level
system with the excited state $\left|e\right\rangle $ ($F=1$) and
the ground state $\left|g\right\rangle $ ($F=0$). The two atoms
are initially excited and the state of the system is given by $\left|e_{1}\right\rangle \otimes\left|e_{2}\right\rangle =\left|e_{1}e_{2}\right\rangle $.
Eventually one of the two atoms spontaneously decays to its ground
state emitting a photon with a wavelength $\lambda$ and energy $\hbar\omega$.
If the interatomic distance $r^{\prime}$ is much smaller than $\lambda$
(i.e., $kr^{\prime}\ll1$) and the two atoms are identical, then one
cannot say which atom has emitted the photon nor which is in a given
state. In the case of the two-level hydrogen atom discussed here,
this decay rate must be related to that of the corresponding magnetic
dipole transition given by (MKS units) 

\begin{equation}
\Gamma=\frac{\mu_{0}k^{3}\left|\left\langle e\right|\hat{\mathbf{M}}\left|g\right\rangle \right|^{2}}{3\hbar\pi}.\label{eq:Gamma}
\end{equation}

\noindent We can furthermore express the state of the system by either
a symmetric $\left|+\right\rangle $ or antisymmetric $\left|-\right\rangle $
combination of the $\left|e_{1}g_{2}\right\rangle $ and $\left|g_{1}e_{2}\right\rangle $
state vectors such as

\begin{eqnarray}
\left|+\right\rangle  & = & \frac{1}{\sqrt{2}}\left(|e_{1}g_{2}\rangle+|g_{1}e_{2}\rangle\right)\label{eq:+state}\\
\left|-\right\rangle  & = & \frac{1}{\sqrt{2}}\left(|e_{1}g_{2}\rangle-|g_{1}e_{2}\rangle\right),\label{eq:-state}
\end{eqnarray}

\noindent which, at this stage of our analysis, have the same energy
and are thus degenerate (see below). 

We now refine this model by adding the magnetic dipole-dipole interaction
term to the system's Hamiltonian. In this model the magnetic dipole
from one atom, say, $\hat{\mathbf{M}}_{1}$, interacts with the magnetic
field $\hat{\mathbf{B}}_{2}$ due to the dipole of the other atom
located at a position $\mathbf{r}^{\prime}=r^{\prime}\mathbf{e}_{r^{\prime}}$
away in the near-field, where $kr^{\prime}\ll1$ \citep{Jackson1999}

\begin{equation}
\hat{{\bf B}}_{2}(\mathbf{r}^{\prime})=\frac{\mu_{0}}{4\pi}\left[\frac{\cos\left(kr^{\prime}\right)}{r^{\prime3}}+\frac{\sin\left(kr^{\prime}\right)}{r^{\prime2}}\right]\left[3\mathbf{e}_{r^{\prime}}\left(\mathbf{e}_{r^{\prime}}\cdot\hat{{\bf M}}_{2}\right)-\hat{{\bf M}}_{2}\right].\label{eq:B(r')}
\end{equation}

\noindent It can be shown that when the two dipoles are aligned, the
term of the interaction Hamiltonian that is relevant to the present
discussion is 
\begin{equation}
\hat{H}_{\mathrm{dd}}=-\frac{\mu_{0}k^{3}\mu_{\mathrm{B}}^{2}}{2\pi}\left(3\left|\beta\right|^{2}-1\right)\left[\frac{\cos\left(kr^{\prime}\right)}{\left(kr^{\prime}\right)^{3}}+\frac{\sin\left(kr^{\prime}\right)}{\left(kr^{\prime}\right)^{2}}\right]\left(\hat{R}_{1}^{+}\hat{R}_{2}^{-}+\hat{R}_{1}^{-}\hat{R}_{2}^{+}\right),\label{eq:Hdd}
\end{equation}
with $\beta=\mathbf{e}_{L}\cdot\mathbf{e}_{r^{\prime}}$. The raising/lowering
operators $\hat{R}_{1}^{+}$, $\hat{R}_{1}^{-}$, etc., are defined
in Equation (\ref{eq:R+-}) and the LCP unit vector state $\mathbf{e}_{L}$
in Equation (\ref{eq:e+}), while $\mu_{\mathrm{B}}$ is the Bohr
magneton. It can further be shown, through a simple diagonalization
exercise, that this interaction Hamiltonian lifts the degeneracy between
the $\left|+\right\rangle $ and $\left|-\right\rangle $ states of
Equations (\ref{eq:+state}) and (\ref{eq:-state}), with their corresponding
energies becoming \citep{Protsenko2006} 

\begin{equation}
E_{\pm}=E_{0}\pm\Delta E,\label{eq:E+-}
\end{equation}

\noindent with $E_{0}$ the unperturbed energy of the states and 

\begin{equation}
\Delta E=\frac{\mu_{0}k^{3}\mu_{\mathrm{B}}^{2}}{2\pi}\left(3\left|\beta\right|^{2}-1\right)\left[\frac{\cos\left(kr^{\prime}\right)}{\left(kr^{\prime}\right)^{3}}+\frac{\sin\left(kr^{\prime}\right)}{\left(kr^{\prime}\right)^{2}}\right].\label{eq:DeltaE}
\end{equation}

\noindent It is then also possible to verify through Equation (\ref{eq:Gamma}),
setting $\hat{\mathbf{M}}=\hat{\mathbf{M}}_{1}+\hat{\mathbf{M}}_{2}$,
that the transition rates $\Gamma_{\pm}$ for the $\left|\pm\right\rangle $
states with the initial $\left|e_{1}e_{2}\right\rangle $ and final
$\left|g_{1}g_{2}\right\rangle $ states are

\begin{eqnarray}
\Gamma_{+} & = & 2\Gamma\label{eq:Gamma+}\\
\Gamma_{-} & = & 0,\label{eq:Gamma-}
\end{eqnarray}

\noindent where $\Gamma$ is the transition rate of a single atom
acting independently. 

The differentiation of energy levels brought about by the magnetic
dipole-dipole interaction is, therefore, seen to be a function of
$\beta$ and the interatomic distance $r^{\prime}$. More important,
however, is the doubling of the transition rate for the symmetric
state $\left|+\right\rangle $, and the cancellation of that of the
antisymmetric state $\left|-\right\rangle $. Such enhanced and reduced
rates are respectively associated to superradiance and subradiance.
This scenario for the two-atom system is depicted in Figure \ref{fig:two-atoms}.

This behavior can also be understood by considering the symmetry of
the system's Hamiltonian and states. The fact that, as could easily
be verified, the Hamiltonian of the system of two atoms (including
the magnetic dipole term $\hat{V}_{\mathrm{MD}}$) is totally symmetric
under the permutation of the two aligned atoms when $kr^{\prime}\ll1$
implies that only states of like symmetry can be coupled. It follows
that since the initial state $\left|e_{1}e_{2}\right\rangle $ of
the fully inverted system is also symmetric, it can only couple to
the $\left|+\right\rangle $ intermediate state, and from there to
the symmetric ground state of the system $\left|g_{1}g_{2}\right\rangle $.
Accordingly, it is interesting to note that under these conditions
a system prepared in the intermediate antisymmetric state $\left|-\right\rangle $
will not decay to the ground state since $\Gamma_{-}=0$. This is
evidently different from the case of a non-coherent system where the
both atoms eventually decay to their individual ground state $\left|g\right\rangle $
at the rate $\Gamma$. We therefore see that superradiance and subradiance
are characteristics of a coherent system, where the intensity of radiation
does not scale linearly with the number of atoms, as is the case for
a non-coherent system.

When the effect discussed here is generalized to a sample composed
of $N$ atoms confined within a volume $\mathcal{V}\ll\lambda^{3}$
(a small-sample), we find that some of the conditions that prevailed
for the two-atom case are not realized. Most importantly, Equations
(\ref{eq:E+-}) and (\ref{eq:DeltaE}) indicate that this interaction
leads to a distribution of energy levels in the system unless the
atoms all have similar nearest neighborhood (e.g., a ring-like periodic
distribution of atoms; \citealt{Gross1982}). This spread in energy
levels will tend to reduce the strength of the superradiance effect. 

It has nonetheless been observed through numerical calculations and
experiments that coherent behaviors still apply to $N$-atom small-sample
systems where radiation is of long enough wavelength \citep{Gross1979},
as is the case for 21 cm line. For $kr^{\prime}\sim1$ the ratio $\Delta E/E_{0}\sim\Gamma/\omega$
is exceedingly small for the 21 cm line, and the time-scale associated
with the energy shifts is on the order of $\hbar/\Delta E\sim\left(kr^{\prime}\right)^{3}\Gamma^{-1}$
\citep{Benedict1996}, which for the H{\footnotesize I} densities
considered in this paper renders this type of dephasing negligible.
As will be discussed later, dephasing due to collisions are more likely
to set the time-scale for homogeneous dephasing. The same is not necessarily
true at short wavelengths where it is very difficult to place a large
number of atoms within a sub-wavelength dimension in a regular pattern,
and in such a sample strong  dipole-dipole interactions break the
symmetry and terminate the coherent behavior by introducing large
energy-level shifts. Thus most of the experimental observations of
superradiance took place at longer wavelengths (i.e., in the infrared
as opposed to optical; \citealt{Benedict1996}). 

For an inverted $N$-atom small-sample with initially uncorrelated
dipoles the first photon emitted by one of the atoms interacts with
the dipole moments of the other atoms, resulting in the build up of
correlation between them. After some time, known as the delay time
$t_{\mathrm{D}}$, a very high degree of correlation is developed
in the system, where in the strongest superradiance regime, the $N$
microscopic dipoles eventually act like one macroscopic dipole. The
rate of emission is then enhanced to $N\Gamma$, while the radiation
intensity is proportional to $N^{2}$ and becomes highly directional,
being focussed in a beam with a temporal half-width on the order of
\textbf{$1/\left(N\Gamma\right)$}. 

It should also be noted that the correlation between dipoles can be
triggered by an external source such as an input radiation field.
This can happen if the input radiation field is stronger than the
spontaneous fluctuations in the sample, and the coupling of the dipoles
to the external field leads to coherent behaviors. An enhancement
of radiation through coupling to an external field is called \textit{triggered
superradiance} \citep{Benedict1996}. 

\begin{figure}[th]
\epsscale{0.7}\plotone{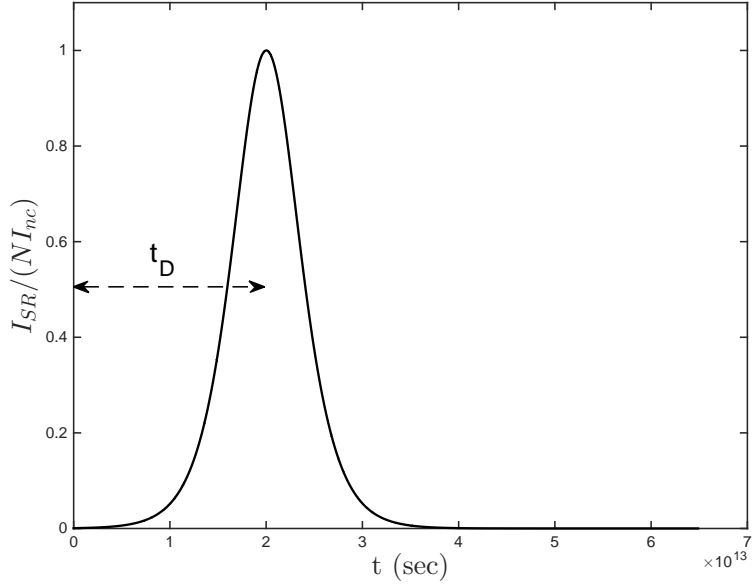}

\caption{\label{fig:ideal-small-sample}The ideal H{\footnotesize I} small-sample
superradiant system. The radiation intensity is plotted as a function
of time $t$, for $N=75$ atoms confined within a cube of $L=4$ cm.
After the delay time $t_{\mathrm{D}}=2.0\times10^{13}$ seconds the
system radiates coherently in a single burst of radiation.}
\end{figure}

It can be shown that the superradiance radiation intensity $I_{\mathrm{SR}}$
of an ideal H{\footnotesize I} small-sample composed of $N$ inverted
atoms is given by \citep{Dicke1954, Gross1982, Benedict1996}

\begin{equation}
I_{\mathrm{SR}}=N^{2}\hbar\omega\Gamma\cosh^{-2}\left[N\Gamma\left(t-t_{\mathrm{D}}\right)\right],\label{eq:I_SR-ideal-small-sample}
\end{equation}

\noindent where $\hbar\omega$ is the energy of the corresponding
atomic transition and the aforementioned delay time $t_{\mathrm{D}}=\left(N\Gamma\right)^{-1}\ln\left(N\right)$.
In Figure \ref{fig:ideal-small-sample} the radiation intensity of
a H{\footnotesize I} small-sample with $N=75$ atoms confined within
a cube of length $4$ cm ($\simeq\lambda/5$ for the 21 cm line) is
plotted as a function of time using Equation (\ref{eq:I_SR-ideal-small-sample}).
The intensity is normalized to $NI_{\mathrm{nc}}$, where $I_{\mathrm{nc}}=N\hbar\omega\Gamma$
is for the corresponding non-coherent small-sample. It can be seen
in Figure \ref{fig:ideal-small-sample} that the energy stored in
the small-sample is radiated away in a single burst. After time $t=t_{\mathrm{D}}$,
the intensity reaches its maximum value, $N$ times that of the non-coherent
intensity, and the peak intensity of the normalized plot becomes equal
to one. In this H{\footnotesize I} sample, $\Gamma^{-1}=3.5\times10^{14}$
sec \citep{Draine2011}, the delay time $t_{\mathrm{D}}=2.0\times10^{13}$
sec, and the characteristic time of superradiance is $T_{\mathrm{R}}=\left(N\Gamma\right)^{-1}=4.6\times10^{12}$
sec. It should also be pointed out that in such a sample the correlation
between dipoles is initiated by internal spontaneous fluctuations
and it is assumed that we are dealing with an ideal system, where
the dipole-dipole symmetry breaking effects are negligible and there
are no other relaxation mechanisms (i.e., cooperative emission is
the only decay mechanism). 

In a real system, there are some relaxation and dephasing effects
that compete with the build-up of the correlation, and in order to
subsequently have superradiance, its characteristics time-scale $T_{\mathrm{R}}$
and delay time\textbf{ $t_{\mathrm{D}}$} must be shorter than (in
some exceptional cases on the order of) the relaxation/dephasing time-scales
\citep{Gross1982, Benedict1996}. The non-ideal case will be discussed
in Section \ref{sub:Non-ideal-Case}.

\subsubsection{Two Hydrogen Atoms Separated by a Larger interatomic Distance $\left(r'>\lambda\right)$\label{sub:Two-Atom-large}}

\begin{figure}[h]
\epsscale{0.45}\plotone{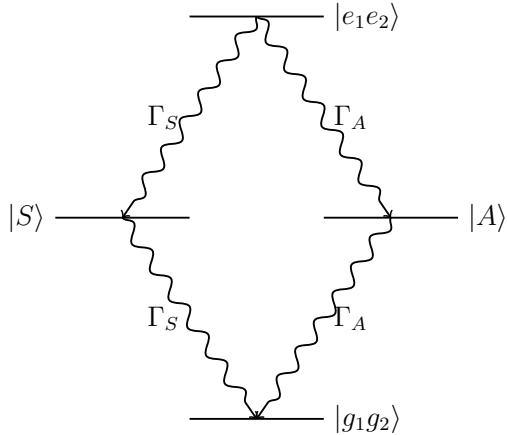}

\caption{\label{fig:two-atoms-large}The two-hydrogen-atom system with $r^{\prime}>\lambda$.
The upper and lower symmetric states $\left|e_{1}e_{2}\right\rangle $
and $\left|g_{1}g_{2}\right\rangle $, respectively, couple to the
intermediate symmetric $\left|S\right\rangle $ and antisymmetric
$\left|A\right\rangle $ states with the corresponding transition
rates $\Gamma_{S}$ and $\Gamma_{A}$. The direction of emission of
the first photon determines the intermediate state of the system,
and the direction of the second photon is correlated with that of
the first.}
\end{figure}

Let us still assume that the atoms are prepared initially in their
excited states, with the state of the two-atom system given by $\left|e_{1}e_{2}\right\rangle $.
Similar to the sub-wavelength case, a first photon is radiated leaving
the system in an intermediate state, which unlike as we did for the
sub-wavelength case, will be described with any combination of $\left|e_{1}g_{2}\right\rangle $
and $\left|g_{1}e_{2}\right\rangle $ states with each having equal
probability contributions, i.e., not only by the $\left|+\right\rangle $
and $\left|-\right\rangle $ states. More precisely, if we associate
the general symmetric state 

\begin{eqnarray}
\left|S\right\rangle  & = & \frac{1}{\sqrt{2}}\left(|e_{1}g_{2}\rangle+e^{i\phi}|g_{1}e_{2}\rangle\right)\label{eq:+state-large-r}
\end{eqnarray}
to the intermediate state shown in the left side of of Figure \ref{fig:two-atoms-large},
then we should assign its orthogonal antisymmetric state 

\begin{eqnarray}
\left|A\right\rangle  & = & \frac{1}{\sqrt{2}}\left(|e_{1}g_{2}\rangle-e^{i\phi}|g_{1}e_{2}\rangle\right)\label{eq:state-large-r}
\end{eqnarray}
to the intermediate state on the right side of the figure \citep{Dicke1964}.
In Equations (\ref{eq:+state-large-r}) and (\ref{eq:state-large-r}),
$\phi$ is a phase term discriminating between the multiple choices
for the intermediate states. To get a better understanding of the
transition probabilities for these states, it is useful to refer to
Equation (\ref{eq:V_MDquantized}) for the magnetic dipole interaction
term with a radiation field for two atoms separated by $\mathbf{r}^{\prime}$
and for a given $\mathbf{k}$. We then find that for coupling to,
say, the $\left|e_{1}e_{2}\right\rangle $ state the following term
comes into play 

\begin{equation}
\hat{V}_{\mathrm{MD}}\propto\hat{R}_{1}^{+}+\hat{R}_{2}^{+}e^{ikr^{\prime}\cos\left(\theta^{\prime}\right)},\label{eq:WMDlarge}
\end{equation}

\noindent where $\theta^{\prime}$ is the angle between $\mathbf{k}$
and $\mathbf{r}^{\prime}$. Given that the transition probability
(and rates) are proportional to $\left|\left\langle e_{1}e_{2}\right|\hat{V}_{\mathrm{MD}}\left|S\right\rangle \right|^{2}$
and $\left|\left\langle e_{1}e_{2}\right|\hat{V}_{\mathrm{MD}}\left|A\right\rangle \right|^{2}$
\citep{Grynberg2010}, we calculate using Equation (\ref{eq:+state-large-r}),
(\ref{eq:state-large-r}), and (\ref{eq:WMDlarge}) 

\begin{eqnarray}
\Gamma_{S} & \propto & \cos^{2}\left\{ \frac{1}{2}\left[\phi-kr^{\prime}\cos\left(\theta^{\prime}\right)\right]\right\} \label{eq:GammaS}\\
\Gamma_{A} & \propto & \sin^{2}\left\{ \frac{1}{2}\left[\phi-kr^{\prime}\cos\left(\theta^{\prime}\right)\right]\right\} .\label{eq:GammaA}
\end{eqnarray}

We therefore see that, although the first photon can be emitted in
any direction $\theta^{\prime}$, its direction of emission determines
$\phi$ and the intermediate state of the system since the transition
probabilities peak at $\phi-kr^{\prime}\cos\left(\theta^{\prime}\right)=2m\pi$
for $\Gamma_{S}$ and $\phi-kr^{\prime}\cos\left(\theta^{\prime}\right)=n\pi$
for $\Gamma_{A}$ ($m$ and $n\neq0$ are integers). Going through
the same exercise for the $\left|g_{1}g_{2}\right\rangle $ state
shows a similar dependency on $\theta^{\prime}$ and $\phi$ as in
Equations (\ref{eq:GammaS}) and (\ref{eq:GammaA}), which implies
that these transitions rates will also be likely to peak at the same
value of $\phi$. It follows that there is an angular correlation
between two successive photons, where the direction of the second
photon is correlated with the direction of the first. This angular
correlation can take place even when the atoms are placed several
wavelengths apart, as a result of their coupling to a common electromagnetic
field, and favors intense radiation along elongated geometries (e.g.,
pencil-like or cylindrical structures; \citealt{Dicke1964}). 

Depending on the intermediate state of the system two different classes
of transitions are possible; coherent and non-coherent. If the emission
of the first photon leaves the system in a symmetric intermediate
state (e.g., $\left|S\right\rangle $) the symmetric coupling to the
radiation field results in the coherent behavior and consequently
the system decays to the symmetric ground state $\left|g_{1}g_{2}\right\rangle $
with the corresponding transition rate shown in Figure \ref{fig:two-atoms-large}.
On the other hand, if the system is in the antisymmetric intermediate
state $\left|A\right\rangle $ the coupling to the radiation field
will be antisymmetric under the exchange of the atoms as they interact
with the non-uniform electromagnetic field, and the system decays
non-coherently to the ground state $\left|g_{1}g_{2}\right\rangle $
with the decay rate $\Gamma_{A}$. 

We can also explain this classically by considering two classical
radiators separated by a distance $r^{\prime}>\lambda$. Over large
distances, the phase and the polarization of the radiation field emitted
by each radiator varies from place to place. When the radiation from
the two identical radiators interferes, the intensity of the total
field is given by

\begin{eqnarray}
I_{\mathrm{tot}} & \varpropto & \left\langle \left({\bf B}_{1}+{\bf B}_{2}\right)^{2}\right\rangle \\
 & \varpropto & \left\langle B_{1}^{2}\right\rangle +\left\langle B_{2}^{2}\right\rangle +2\left\langle {\bf B}_{1}\cdot{\bf B}_{2}\right\rangle ,\label{eq:classical-interference-intensity}
\end{eqnarray}
and can become as large as four times of the intensity of a single
radiator $I_{0}$ if

\begin{eqnarray}
\left\langle {\bf B}_{1}\cdot{\bf B}_{2}\right\rangle  & \sim & I_{0}.\label{eq:perfect-correlation}
\end{eqnarray}
If the phase of the radiation from different radiators does not match
perfectly, the term containing the correlation $\left\langle {\bf B}_{1}\cdot{\bf B}_{2}\right\rangle $
in Equation (\ref{eq:classical-interference-intensity}) becomes smaller
than $I_{0}$ and consequently the total intensity $I_{\mathrm{tot}}$
decreases until it reaches its minimum for completely out of phase
radiators. Furthermore, the correlation term can vanish when radiators
act independently. In this case, the total intensity becomes equal
to the sum of the intensities of the two independent radiators (the
so-called non-coherent system).

\subsection{The $N$-atom Large-sample $\left(\mathcal{V}>\lambda^{3}\right)$\label{sub:Large-N-atom-Sample} }

We can extend our discussion for the case of two distant atoms (i.e.,
$r^{\prime}>\lambda$) to a large-sample consisting of $N$ atoms
distributed over a volume $\mathcal{V}>\lambda^{3}$. As stated above,
the build-up of correlations in an extended $N$-atom sample can be
understood as a constructive interference of the radiation by different
atoms. In a large-sample as a result of propagation over large distances
(i.e., larger than $\lambda$) the phase of the radiation varies throughout
the sample ($kr^{\prime}\gg1$ and $e^{i\mathbf{k\cdot}\mathbf{r^{\prime}}}\nsim1$).
Consequently, the phase of the atomic magnetization differs with position.
In an inverted large-sample, the radiation from different atoms interfere
with each other, and when the magnetization of the radiators are perfectly
in phase, an intense propagating wave is produced in one direction
(the phase-matching condition cannot be satisfied in all directions). 

In order to better understand the phase-matching process, it is useful
to go back to the angular correlation effect described in the two-atom
case. In a large-sample of $N$ inverted atoms, when the first photon
is emitted other atoms interact with its radiation field and the direction
of the next photon is affected by the first one. In a more general
sense, when a photon is radiated in a particular direction ${\bf k}$,
it becomes more probable to observe the second photon in the same
direction ${\bf k}$ than any other direction. Thus, as the atoms
radiate, an angular correlation builds up in the sample that triggers
the phase-matching process in a well-defined direction. 

Ideal superradiance is the result of the symmetrical evolution of
an atom-field system, and in a large-sample, the propagation effects
result in the non-uniform evolution of the atoms in the sample. In
order to better understand propagation effects in a large-sample,
the atomic medium can be divided into small identical slices with
dimensions larger than $\lambda$ but much smaller than the length
of the sample. A microscopic dipole is then associated to each slice
with its magnitude being proportional to the number of excited atoms
in the corresponding slice. At the beginning, the dipoles in different
slices are independent and their radiation uncorrelated. After some
time (or the so-called retarded-time delay $\tau_{\mathrm{D}}$; see
Equation (\ref{eq:delay-time-large-sample}) below), as they interact
with their common radiation field, the dipoles lock to a common phase,
and act as a single macroscopic dipole radiating intensely with $I_{\mathrm{SR}}=NfI_{\mathrm{nc}}$,
where $Nf$ is the enhancement factor of the superradiant intensity
$I_{\mathrm{SR}}$ over the non-coherent intensity $I_{\mathrm{nc}}$
determined by the efficiency of the common phase-locking process (through
the value of $f\leq1$). The enhancement factor $Nf$ can become very
large in samples with $N\gg1$, and it converges to $N$ in an atomic
system with dimensions of the order of $\lambda$ resulting in $I_{\mathrm{SR}}=NI_{\mathrm{nc}}$
for the most efficient phase-locking process seen in a small-sample.
In other words, $f<1$ implies a limited coherent behavior in a large-sample
resulting in a smaller output intensity and weakened superradiance,
whereas $f=1$ indicates a fully coherent behavior leading to an intense
radiation and perfect superradiance \citep{Gross1982, MacGillivray1976}.\textcolor{blue}{{} }

This approach has the shortcoming that it cannot explain the initiation
of the radiation in the system by spontaneous fluctuations, and to
overcome this problem phenomenological fluctuations of dipoles in
the initial stages of the evolution can be added to the formalism.
In contrast, triggered superradiance can be fully explained in this
manner as the correlation process is initiated by an external field,
which can be defined classically. It must be pointed out that the
results of this method are valid only if the propagation time of the
radiation $\tau_{\mathrm{E}}$ through a sample of length $L$ (i.e.,
$\tau_{\mathrm{E}}=L/c$) is smaller than the superradiance characteristic
time $T_{\mathrm{R}}$ given by 

\begin{equation}
T_{\mathrm{R}}=\tau_{\mathrm{sp}}\frac{16\pi}{3n\lambda^{2}L},\label{eq:TR-largesample}
\end{equation}

\noindent where $\tau_{\mathrm{sp}}=1/\Gamma$ is the spontaneous
decay time of a single atom and $n$ the density of inverted atoms
(see \citealt{MacGillivray1976, Rosenberg1981}, and Appendix \ref{sec:Non-linear-regime}).
This condition (i.e., $\tau_{\mathrm{E}}<T_{\mathrm{R}}$) is known
as Arecchi-Courtens condition, and it ensures that the atomic magnetization
in different parts of the sample can lock into a common phase and
coherent behavior can develop through the sample. 

In Appendix \ref{sec:Theoretical-Model}, we derive the evolution
equations for the radiation field and the atomic system using the
Heisenberg representation, while in Appendix \ref{sec:Non-linear-regime}
we solve the corresponding Maxwell-Bloch system of equations, at resonance,
within the framework of the slowly varying envelope approximation
(SVEA). To do so, we adopted the following form for the radiation
magnetic field and atomic magnetization

\begin{eqnarray}
\hat{B}_{L}^{\pm}\left(\mathbf{r},t\right) & = & \hat{B}_{0}^{\pm}\left(\mathbf{r},t\right)e^{\pm i\left(kz-\omega t\right)}\label{eq:BL-SVEA}\\
\hat{\mathcal{M}}^{\pm}\left(\mathbf{r},t\right) & = & \hat{\mathcal{M}}_{0}^{\pm}\left(\mathbf{r},t\right)e^{\pm i\left(kz-\omega t\right)},\label{eq:M-SVEA}
\end{eqnarray}

\noindent with $\hat{B}_{0}^{\pm}$ and $\hat{\mathcal{M}}_{0}^{\pm}$
corresponding slowly varying envelope operators. The superradiance
of a cylindrical large-sample of length $L$ under ideal conditions
is then found to be determined by the following equations for, respectively,
the magnetization, the population inversion, and the magnetic field

\begin{eqnarray}
\hat{\mathcal{M}}_{0}^{+} & = & \frac{\mu_{\mathrm{B}}N}{2\sqrt{2}V}\sin\left(\theta\right)\label{eq:M0+(theta)}\\
\hat{\mathbb{N}} & = & \frac{N}{V}\cos\left(\theta\right)\label{eq:N(theta)}\\
\hat{B}_{0}^{+} & = & \frac{i\mu_{\mathrm{B}}}{2\sqrt{2}\gamma}\frac{\partial\theta}{\partial\tau},\label{eq:B0+(theta)}
\end{eqnarray}

\noindent where $\gamma=\mu_{\mathrm{B}}^{2}/2\hbar$. The solution
for the Bloch angle $\theta$ as a function of the retarded-time $\tau=t-L/c$
is obtained through the so-called Sine-Gordon equation 

\begin{equation}
\frac{d^{2}\theta}{dq^{2}}+\frac{1}{q}\frac{d\theta}{dq}=\sin\left(\theta\right)\label{eq:S-G}
\end{equation}

\noindent with

\begin{equation}
q=2\sqrt{\frac{z\tau}{LT_{\mathrm{R}}}}.\label{eq:q-main_text}
\end{equation}

\begin{figure}[th]
\epsscale{0.7}\plotone{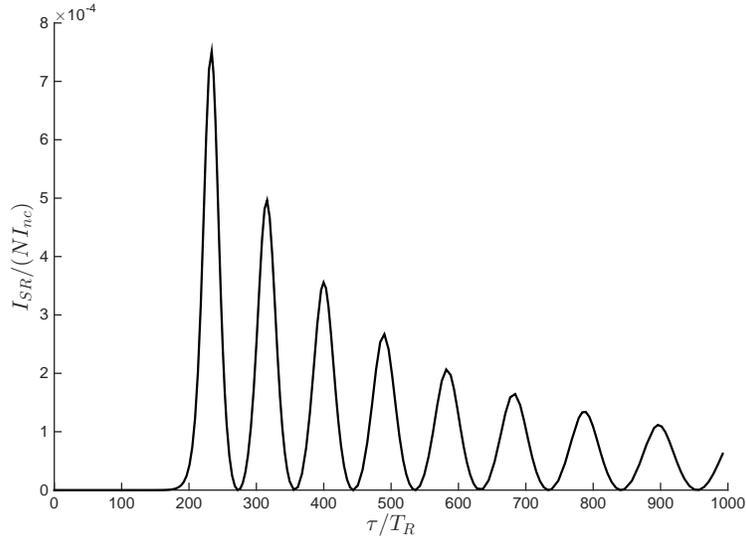}

\caption{\label{fig:idealcase-largesample}The ideal cylindrical H{\footnotesize I}
large-sample. The radiation intensity, scaled to $NI_{\mathrm{nc}}$,
is plotted versus the retarded time $\tau=t-L/c$ normalized to the
superradiance characteristic time-scale $T_{\mathrm{R}}$. The length
and radius of the cylinder are, respectively, $L=0.02\,cT_{\mathrm{R}}$
and $w=0.036\left(cT_{\mathrm{R}}\right)^{1/2}$.}
\end{figure}

In Figure \ref{fig:idealcase-largesample} we show the solution for
the radiation intensity of such an ideal cylindrical large-sample
of H{\footnotesize I} atoms of length $L=0.02\,cT_{\mathrm{R}}$,
where $c$ is the speed of the light, by numerically solving Equations
(\ref{eq:M0+(theta)}) to (\ref{eq:q-main_text}). For these calculations
we set a Fresnel number of unity to reduce the impact of diffraction
losses, which are not taken into account in our model. This yields
a cylinder of radius 

\begin{equation}
w=\sqrt{\frac{\lambda L}{\pi}},\label{eq:w}
\end{equation}

\noindent which for our ideal sample results in $w=0.036\left(cT_{\mathrm{R}}\right)^{1/2}$.

In this sample, the Arecchi-Courtens condition is satisfied (i.e.,
$\tau_{\mathrm{\mathrm{E}}}\ll T_{\mathrm{R}}$) allowing the use
of the homogeneous condition $\theta_{0}=4.9\times10^{-12}$ rad for
the initial value of the Bloch angle. More precisely, for the large-sample
used for the figure we assumed that internal fluctuations dominate
over triggered superradiance, and the initial Bloch angle was set
with $\theta_{0}=2/\sqrt{N}$ \citep{Gross1982}.

In Figure \ref{fig:idealcase-largesample}, the retarded-time axis
is scaled to $T_{\mathrm{R}}$ and the radiation intensity to $NI_{\mathrm{nc}}$,
i.e., the number of inverted atoms times the corresponding non-coherent
intensity that would otherwise be expected from such a sample. More
precisely, for comparison purposes we consider the non-coherent intensity
emanating through the sample's end-fire (i.e., the end facing the
observer) of area $A$, within the superradiance radiation beam solid-angle
$\phi_{\mathrm{D}}=\lambda^{2}/A$ (in the direction $\mathbf{k}$
along which the phase-locking condition is satisfied) normalized to
the solid angle associated to the total non-coherent radiation. We
thus have

\begin{eqnarray}
I_{\mathrm{nc}} & = & N\hbar\omega\left(\frac{1}{A\tau_{\mathrm{sp}}}\right)\left(\frac{\phi_{\mathrm{D}}}{4\pi}\right)\nonumber \\
 & = & \frac{4}{3}\frac{\hbar\omega}{AT_{\mathrm{R}}},\label{eq:Inc-largesample}
\end{eqnarray}

\noindent where Equation (\ref{eq:TR-largesample}) was used for the
last step and $N\hbar\omega$ is the total energy initially stored
in the sample. As shown in the figure, this energy is radiated away
through multiple bursts, a phenomenon known as ringing effect. This
effect can be explained by the fact that atoms in different locations
in the sample radiate at different times. In other words, an atom
at location $z=z_{0}$, prepared in the excited state at $\tau=0$,
radiates its energy away and decays to its ground state, then later
on absorbs the energy radiated by another atom at a location $z<z_{0}$
and becomes excited leading to another radiation event. 

In a large-sample, just as in a small-sample, internal field fluctuations
or an external field trigger superradiance, and after the delay time
$\tau_{\mathrm{D}}$ the atoms radiate coherently. But contrary to
a small-sample, the large-sample delay time depends on the initial
conditions and is given by \citep{Benedict1996}

\begin{eqnarray}
\tau_{\mathrm{D}} & \simeq & \frac{T_{\mathrm{R}}}{4}\left|\ln\left(\frac{\theta_{0}}{2\pi}\right)\right|^{2}.\label{eq:delay-time-large-sample}
\end{eqnarray}
In Fig\textcolor{black}{ure \ref{fig:idealcase-largesample} th}e
first burst of radiation occurs after $\tau_{\mathrm{D}}\simeq160\,T_{\mathrm{R}}$,
which is consistent with the value one finds using Equation (\ref{eq:delay-time-large-sample}).
As can also be seen, this first intensity burst only carries out a
fraction of the total energy stored in the sample, while the remaining
radiation happens through subsequent bursts. The number of burst events
depends on the length of the sample, and as the length is increased
radiation emanates through a larger number of bursts, while the peak
intensity of consecutive burst events gradually drops. This is a consequence
of energy conservation and the fact that in a larger (i.e., longer)
sample radiation from different groups of atoms along the sample arrive
at the end-fire at different times, and the process of absorbing the
radiation, developing correlations between the dipoles, and eventually
re-emitting the radiation repeats multiple times over a very long
period of time. On the other hand, when the length of the sample is
decreased the ringing effect becomes weaker until, for a small-sample
of dimension of order of $\lambda$, it totally washes out and we
only observe a single burst of radiation carrying away all the energy
stored in the system (as in Figure \ref{fig:ideal-small-sample}).
Finally, we note that although the maximum radiation intensity seen
in Figure \ref{fig:idealcase-largesample} seem to imply that $f\sim0.001\ll1$,
the large number of atoms present in the sample ensures that $I_{\mathrm{SR}}\gg I_{\mathrm{nc}}$
(see Section \ref{sec:Discussion}). We should also note, however,
that the Sine-Gordon equation is very sensitive to initial conditions.
It therefore follows that the exact shape of the intensity curve,
e.g., the number of bursts in Figure \ref{fig:idealcase-largesample},
is also strongly dependent on $\theta_{0}$.

\subsubsection{Non-ideal Case -- Dephasing Effects\label{sub:Non-ideal-Case}}

As was mentioned earlier, the characteristics time-scale of superradiance
$T_{\mathrm{R}}$ and the delay time $\tau_{\mathrm{D}}$ (for a large-sample)
must be shorter than the relaxation/dephasing time-scales to allow
the build-up of correlations in a non-ideal sample. These effects
include natural broadening due to the spontaneous decay time-scale
$\tau_{\mathrm{sp}}$ of a single atom and collisional broadening
related to the mean time between collisions $\tau_{\mathrm{coll}}$
for an atom in the sample. Although, as was stated in Section \ref{sec:Introduction},
our analysis is aimed at regions of the ISM where thermal equilibrium
has not been reached and where consequently the assignation of a temperature
to determine, for example, collision rates is perhaps ill-defined,
we will nonetheless adopt such a procedure for the rest of our discussion
to get a sense of the time-scales involved. Accordingly, in a H{\footnotesize I}
gas different types of collisions can take place depending on the
temperature and density. For environments of temperatures ranging
from approximately $10\,\mathrm{K}$ to $300\,\mathrm{K}$, which
are the focus of our analysis, collisions between two neutral hydrogen
atoms (H-H collisions) dominate and fall into two categories: elastic
and inelastic. During an elastic H-H collision, the spacing between
the atomic energy levels are slightly affected but no transition between
them is induced. The change in energy spacing occurs as a result of
short-range interaction forces between the two colliding particles
and induces a phase shift in the wave function of the scattered atoms.
After a number of elastic collisions, an atom can lose coherence with
the interacting radiation field as a result of the randomness in the
perturbations. On the other hand, in an inelastic H-H collision the
internal energy of the hydrogen atoms will be changed. This occurs
when the two hydrogen atoms with oppositely directed electron spins
approach each other at distances less than approximately $10^{-8}$
cm. This process is known as electron exchange or spin de-excitation
effect. As a result of such a collision the induced phase shift can
lead to a change in the internal spin states \citep{Wittke1956},
and it is found that spin de-excitation is the dominating relaxation
process in a high density collision-dominated H{\footnotesize I}
gas. We therefore find that H-H collisions not only can affect the
strength of a potential coherent 21-cm radiation by removing from
the population of the excited hyperfine states, they also contribute
to the line breadth and can change the shape of the spectral line
by affecting the spacing between internal energy levels. For example,
the time-scale of H-H collisions are estimated to be on the order
of $10^{8}$ sec in the case of elastic scattering and $10^{9}$ sec
for spin de-excitations, using the mean effective collisional cross
sections given in \citet{Irwin2007} for a H{\footnotesize I} gas
at $T=100\,\mathrm{K}$ and $n=10\,\mathrm{cm}^{-3}$. The mean time
between collision $\tau_{\mathrm{coll}}$ is thus set to the shortest
of these time-scales and must at least be larger than $T_{\mathrm{R}}$
and $\tau_{\mathrm{D}}$ to allow coherent behavior (see Section \ref{sec:Discussion}).

In addition, other broadening effects, such as Doppler broadening,
are further dephasing mechanisms that can destroy cooperative behavior
if their time-scales (importantly the so-called Doppler dephasing
time, i.e., the reciprocal of the Doppler width) are smaller than
$T_{\mathrm{R}}$ and $\tau_{\mathrm{D}}$ \citep{Meziane2002,Bonifacio1975}.
In a thermally relaxed gas, thermal motions are probably the most
important dephasing effects and result in line broadenings that correspond
to very short dephasing time-scales (e.g., $T_{\mathrm{therm}}\sim10^{-3}$
sec at $T=100$ K). In the presence of such strong dephasing effects,
correlations cannot develop between the dipoles and any coherent interaction
will be terminated right from the start. Hence our earlier comment
that we do not expect to find superradiance under conditions of thermal
equilibrium, but potentially only in (out-of-equilibrium) regions
where strong velocity coherence can be maintained along the line-of-sight.
Furthermore, this condition may be only met among a group of atoms
in such regions, therefore reducing the number of inverted atoms in
the sample that could participate in coherent interactions. However,
we know from maser observations that a high level of velocity coherence
can be achieved in some regions of the ISM, and we expect that superradiance
could happen under similar conditions. As was mentioned in Section
\ref{sec:Introduction}, the main inversion pumping mechanism likely
involved for the 21 cm transition points to the surroundings of H{\footnotesize II}
as potentials sites for superradiance in this spectral line.\textbf{
}It follows that we should also anticipate analogous (i.e., very small)
volume filling factor for the emitting regions of superradiant sources
as for masers. 

\begin{figure}[th]
\epsscale{0.7}\plotone{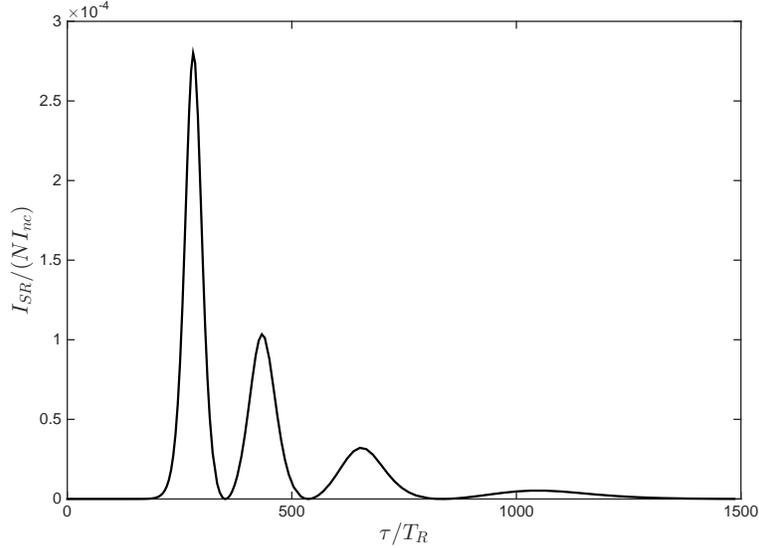}\caption{\label{fig:equaldeaphsing-largesample}The non-ideal H{\footnotesize I}
cylindrical large-sample. All parameters are as in Figure \ref{fig:idealcase-largesample},
except that dephasing/relaxation effects are included for the special
case where they are characterized by a single time-scale $T^{\prime}=541\,T_{\mathrm{R}}$.
The ringing effect is weakened as a result of the dephasing.}
\end{figure}

In Figure \ref{fig:equaldeaphsing-largesample} the intensity of the
H{\footnotesize I} large-sample discussed in Section \ref{sub:Large-N-atom-Sample}
(and presented in Figure \ref{fig:idealcase-largesample} for the
ideal case) is plotted as a function of the retarded-time $\tau$
for the special case where dephasing/relaxation effects are included
and characterized by a single time-scale set to $T^{\prime}=541\,T_{\mathrm{R}}$
(see Sections \ref{sec:Discussion} and Appendix \ref{sec:Non-linear-regime}).
These results were obtained by once again numerically solving Equation
(\ref{eq:S-G}), while the magnetization and population inversion
are given by

\begin{eqnarray}
\hat{\mathcal{M}}_{0}^{+} & = & \frac{\mu_{\mathrm{B}}N}{2\sqrt{2}V}\sin\left(\theta\right)e^{-\tau/T^{\prime}}\label{eq:M0+-dephasing}\\
\hat{\mathbb{N}} & = & \frac{N}{V}\cos\left(\theta\right)e^{-\tau/T^{\prime}},\label{eq:N-dephasing}
\end{eqnarray}

\noindent and the dimensionless parameter

\begin{equation}
q=2\sqrt{\frac{z\tau^{\prime}}{LT_{\mathrm{R}}}}\label{eq:q-tau'}
\end{equation}

\noindent with 

\begin{equation}
\tau^{\prime}=T^{\prime}\left(1-e^{-\tau/T^{\prime}}\right).\label{eq:tau'}
\end{equation}

\noindent The magnetic field is once again given by Equation (\ref{eq:B0+(theta)}).
The intensity and time axes are scaled as those of Figure \ref{fig:idealcase-largesample}
for the ideal H{\footnotesize I} sample. We can see from Figure \ref{fig:equaldeaphsing-largesample}
that the ringing effect seen in the ideal sample is also present here
but is weakened by the dephasing and basically terminated after $\tau\sim1000\,T_{\mathrm{R}}$
(i.e., approximately the dephasing time-scale). The dephasing effects
also affect the maximum energy radiated away through each burst event,
and result in slightly weaker intensities.

\section{Discussion - Cooperative Behavior in a H{\footnotesize I} Gas in
the ISM \label{sec:Discussion}}

As was made evident from our previous discussions, the characteristic
time-scale of superradiance $T_{\mathrm{R}}$ is a fundamental criterion
to consider in the investigation of this cooperative behavior. For
an ideal small-sample of volume $\mathcal{V}<\lambda^{3}\sim10^{4}$
cm$^{3}$, for total hydrogen densities $1$ cm$^{-3}<n_{\mathrm{H}}<100$
cm$^{-3}$ and a population inversion $\eta\sim0.01$, the superradiance
time-scale 

\begin{eqnarray}
T_{\mathrm{R}} & = & \frac{\tau_{\mathrm{sp}}}{\eta n_{\mathrm{H}}\mathcal{V}},\label{eq:tauc}\\
 & \sim & \frac{10^{12}}{n_{\mathrm{H}}}\:\mathrm{s}
\end{eqnarray}

\noindent is such that $T_{\mathrm{R}}\sim10^{10}-10^{12}$ sec, where
$\tau_{\mathrm{sp}}\sim10^{14}$ sec for the 21 cm line. As mentioned
in Section \ref{sub:Non-ideal-Case}, $t_{\mathrm{D}}$ ($\approx T_{\mathrm{R}}$,
in this case) must be shorter than the relaxation/dephasing time-scales
in order to allow the build-up of cooperative behaviors in the sample.
In a typical H{\footnotesize I} gas, an important relaxation mechanism
for an atom is collisional de-excitation, the rate for which is given
by 
\begin{equation}
\tau_{\mathrm{coll}}=\frac{1}{n_{\mathrm{H}}\kappa_{10}}.\label{eq:tcoll}
\end{equation}

\noindent Values for the collisional de-excitation rate coefficient
$\kappa_{10}$ for the hyperfine state $F=1$ over a range of temperatures
can be found in \citet{Zygelman2005}, and for $10\:\mathrm{K}<T<300\:\mathrm{K}$
we find $\tau_{\mathrm{coll}}\sim10^{8}-10^{12}$ sec for the previous
range of atomic hydrogen densities. These collision rates become $\tau_{\mathrm{coll}}\sim10^{7}-10^{12}$
sec when elastic collisions are considered \citep{Irwin2007}. It
can thus be seen that the expected collisional relaxation time-scales
are likely to render superradiance improbable for small H{\footnotesize I}
sample, especially at higher temperatures. This suggests the investigation
of larger samples containing more atoms. 

Although in a large-sample the general cooperative behavior is weaker
than the ideal Dicke superradiance, the greater number of atoms will
reduce $T_{\mathrm{R}}$, increase the radiation intensity, and make
it more likely to observe the effect. However, it is also important
to realize that not only must the condition $T_{\mathrm{R}}<T^{\prime}$
be realized for establishing superradiance, such that the non-coherent
de-excitation does not become the dominant mechanism to release the
energy of the system, but we must also ensure that $\tau_{\mathrm{D}}<T^{\prime}$
is verified since $\tau_{\mathrm{D}}>T_{\mathrm{R}}$ from Equation
(\ref{eq:delay-time-large-sample}). In other words, the delay time
needed to establish coherence in the sample must also be smaller than
the dephasing time-scale. 

It is important to note that the requirement $\tau_{\mathrm{D}}<T^{\prime}$
effectively sets a threshold that must be met for the onset of superradiance.
From the dependency of $\tau_{\mathrm{D}}$ on the different parameters
(see Equation {[}\ref{eq:delay-time-large-sample}{]}) we find that,
for a given transition, it can only be reduced below $T^{\prime}$
through a corresponding increase of the inverted population's column
density $nL$. It follows that superradiance will only be triggered
when the column density meets or exceeds some critical value \citep{Rajabi2016a}.
Contrary to what is the case for laboratory superradiance experiments
where short laser pulses are used to create the necessary population
inversion, the existence of a threshold also implies that there is
no requirement for the presence of a pulse to initiate superradiance
in the ISM (but see below). It only matters that a critical level
of inversion is reached, and the rate at which it is attained is irrelevant. 

However, stronger superradiance bursts can be achieved in the presence
of po\-pu\-lation-inverting pulses that bring the column density
to levels significantly exceeding its critical value. But in such
cases the values attained for $nL$ can be limited by the pumping
time $T_{\mathrm{P}}$ over which the population inversion is achieved
throughout the sample. For example,\textbf{ }as we increase the length
of the sample $T_{\mathrm{R}}$ and $\tau_{\mathrm{D}}$ decrease,
but on the other hand, it becomes necessary to achieve the population
inversion over a larger length-scale. There are two types of pumping
mechanisms available to achieve the population inversion in an atomic
system: swept pumping and instantaneous pumping \citep{MacGillivray1981,Gross1982}.
In the swept pumping scenario the atomic sample is pumped longitudinally
by a pulse traveling along the sample, and the pumping process is
characterized by a finite pumping time $T_{\mathrm{P}}$, whereas
for instantaneous pumping a transverse excitation causes the simultaneous
excitation of all atoms in the sample resulting in $T_{\mathrm{P}}\approx0$.
When the pumping process cannot be achieved instantaneously, but is
realized over a finite time, the output intensity of a small- or a
large-sample are only affected slightly as long as $T_{\mathrm{P}}<\tau_{\mathrm{D}}$.
The main effect of the finite pumping time is then an increase in
the delay time of the superradiant process in comparison to what one
expects from Equation (\ref{eq:delay-time-large-sample}), i.e., the
actual delay time $\tau_{\mathrm{D}}^{\prime}$ will be longer than
the theoretical delay time $\tau_{\mathrm{D}}$ \citep{MacGillivray1981}. 

However, if pumping occurs\textbf{ }at an approximately\textbf{ }constant
rate\textbf{ }over some pulse\textbf{ }time interval in a H{\footnotesize I}
gas (i.e., $dn(t)/dt\simeq\Lambda$, where $\Lambda$ is a constant)
and the first burst of superradiance radiation is emitted before the
expected inversion density $n$ is achieved throughout the sample,
then we need to replace $n$ by its effective value at time $t=\tau_{\mathrm{D}}^{\prime}$
(i.e., $n_{\mathrm{eff}}=n\left(\tau_{\mathrm{D}}^{\prime}\right)$).
In this case, only a fraction of atoms contribute to the first burst,
and the system cannot emit as much of the energy stored within it
through a series of coherent bursts \citep{MacGillivray1981}. If
the pulse (i.e., pumping) time becomes too long and the first superradiant
burst is emitted before the inversion is achieved along the sample,
only a few coherent bursts can be observed. This happens as the system
reaches a quasi-steady state, in which the growth and depletion of
population occur with the same rates. In the quasi-steady state the
radiation intensity is given by $I=\hbar\omega L\left(dn/dt\right)$,
and is no longer proportional to $N^{2}$ \citep{MacGillivray1981}.
The transition from a superradiant to the quasi-steady regime was
first observed in the laboratory by \citet{Gross1976}. For an astronomical
system, this would correspond to passing from a superradiant system
to an astronomical (i.e., mirror-less) maser. We therefore conclude
that in the ISM superradiance will not happen in a steady state regime,
but will rather be characterized by strong variability in radiation
intensities over time. One could, for example, conceive of an emitting
region harboring a maser that would be episodically modulated with
strong bursts of radiation due to superradiance, perhaps resulting
from some radiative trigger or a sudden decrease in $\tau_{\mathrm{D}}$
(from a corresponding increase in the inverted population; for example,
see\textbf{ }\citealt{Rajabi2016a, Rajabi2016b}).

In a more general context, i.e., without limiting the discussion to
the $21$ cm line, it is also important to note that superradiance
triggered through population-inverting pulses (that bring $nL$ significantly
above its critical value) cannot result if these pulses are due to
collisions alone. This is because of the undesirable consequences
collisions have on the dephasing/relaxation of a sample. That is,
if $T_{\mathrm{P}}$ is the pumping time due to collisions, then we
know from our previous discussion that $T_{\mathrm{P}}<\tau_{\mathrm{D}}$
for pulse-initiated superradiance to be possible. But since in this
case the time-scale for collision dephasing is $T^{\prime}=T_{\mathrm{P}}$
it follows that $\tau_{\mathrm{D}}>T^{\prime}$ and superradiance
will be inhibited by collisions.

When $N\gg1$ we have for the average delay time $\left\langle \tau_{\mathrm{D}}\right\rangle =T_{\mathrm{R}}\ln\left(N\right)$
\citep{Gross1982}, which means that $\left\langle \tau_{\mathrm{D}}\right\rangle $
is usually an order of magnitude or two larger than $T_{\mathrm{R}}$
for the large-samples to be studied. As we will now see, for the range
of densities and temperatures considered for our analysis $\left\langle \tau_{\mathrm{D}}\right\rangle <T^{\prime}$
can be realized in a large set of conditions. In a H{\footnotesize I}
large-sample where all the necessary conditions for superradiance
are fulfilled (i.e., $T_{\mathrm{P}},\tau_{\mathrm{E}}<\left\langle \tau_{\mathrm{D}}\right\rangle <T^{\prime}$,
with sufficient velocity coherence), we can estimate the time-scale
of potential superradiance bursts using the results of our numerical
analyses for the corresponding large-sample, as long as the Fresnel
number is kept close to unity. For the following examples we considered
a density $n_{\mathrm{H}}=10\;\mathrm{cm^{-3}}$ with an inversion
factor $\eta=0.01$, implying an inverted population of 940 atoms
in a volume $\lambda^{3}$. Our results indicate that radiation bursts
over time-scales on the order of days (i.e., from Figure \ref{fig:equaldeaphsing-largesample}
a few hundred times $T_{\mathrm{R}}=10^{3}$ sec, while $\left\langle \tau_{\mathrm{D}}\right\rangle =5.2\times10^{4}$
sec) can be associated to cylindrical H{\footnotesize I} samples
of length and radius $L\approx10^{11}$ cm and $w\approx9\times10^{5}$
cm, respectively, while bursts over time-scales on the order of minutes
(i.e., $T_{\mathrm{R}}=1$ sec and\textbf{ $\left\langle \tau_{\mathrm{D}}\right\rangle =66$
}sec) to samples with $L\approx10^{14}$ cm \textbf{(}approximately
equal to $6\times10^{-5}$ pc; see \citealt{Storer1968}) and $w\approx3\times10^{7}$
cm. In all cases, we have $T^{\prime}\approx\tau_{\mathrm{coll}}>\left\langle \tau_{\mathrm{D}}\right\rangle $
over a wide range of conditions, ensuring that dephasing effects will
not destroy atomic coherent behaviors, and we found $f\approx10^{-4}$
(from Figure \ref{fig:equaldeaphsing-largesample}) with an efficiency
factor $Nf$ ranging from approximately $10^{12}$ to $10^{24}$,
from the shortest to the longest sample length $L$. These results
imply a corresponding amplification factor of $10^{10}$ to $10^{22}$
over the corresponding non-coherent intensity of such samples (taking
into account the non-inverted population). 

Although the samples considered above would probably not yield strong
detections (e.g., for the sample of length $L\approx10^{14}$ cm and
$w\approx9\times10^{7}$ cm we calculate an integrated flux $\sim10^{-22}\,\mathrm{erg}\,\mathrm{s}^{-1}\,\mathrm{cm}^{-2}$
at a distance of $400$ pc), given the small radii considered here
it is unlikely that only a single superradiant system would be realized
in a region harboring an inverted population. That is, if we assume
a reasonable maser spot size for the population inverted region (e.g.,
$w_{\mathrm{spot}}\sim1$ AU), then it becomes possible that a very
large number of superradiant system could simultaneously erupt ($w_{\mathrm{spot}}/w\sim10^{6}$)
and render a strong detection more likely, when the conditions for
superradiance are met \citep{Rajabi2016a}. This leads us to suggest
that despite the simplicity of, and the approximations used in, our
model significant intensity variability due to superradiance could
be detectable for the 21 cm line in some regions of the ISM.

\section{Conclusion\label{sec:Conclusion}}

We have applied the concept of superradiance introduced by \citet{Dicke1954}
to astrophysics by extending the corresponding analysis to the magnetic
dipole interaction characterizing the atomic hydrogen 21 cm line.
Although it is unlikely that superradiance could take place in thermally
relaxed regions and that the lack of observational evidence of masers
for this transition reduces the probability of detecting superradiance,
in situations where the necessary conditions are met (i.e., close
atomic spacing, high\textbf{ }velocity coherence, population inversion,
and long dephasing time-scales compared to the\textbf{ }those related
to coherent behavior), our results suggest that relatively low levels
of population inversion over short astronomical length-scales (e.g.,
as compared to those required for maser amplification) can lead to
the cooperative behavior required for superradiance in the ISM. Given
the results of our analysis, we expect the observational properties
of 21-cm superradiance to be characterized by the emission of high
intensity, spatially compact, burst-like features potentially taking
place over periods ranging from minutes to days. 

As this first paper on this topic has, in part, served as an introduction
to superradiance in astrophysics, much remains to be done. For example,
we have not attempted to characterize the shapes of superradiant spectral
lines or their polarization properties, which for the 21 cm line would
necessitate the consideration of all hyperfine $F=1\leftrightarrow0$
transitions. We thus intend to extend our analysis in subsequent publications
to tackle these questions and investigate superradiance in other important
astronomical spectral lines (e.g., the OH 1612-MHz, CH$_{3}$OH 6.7-GHz,
and H$_{2}$O 22-GHz maser transitions) where observational evidence
for superradiance can be found in the literature \citep{Rajabi2016a, Rajabi2016b}.
It would further be beneficial to broaden the scope of our analysis
to include a wider range of conditions such effects as pumping, diffraction
losses, and different sample geometries associated to different Fresnel
numbers.

\acknowledgements{We thank M. Harwit for bringing this research topic to our attention,
and J. Zmuidzinas for helpful discussions. We are also grateful to
T. Troland for alerting us to the evidence for a 21-cm population
inversion in the Orion Veil. M.H.'s research is funded through the
NSERC Discovery Grant and the Western Strategic Support.}

\appendix{}

\section{Theoretical Model\label{sec:Theoretical-Model}}

\subsection{The Hamiltonian and the Maxwell-Bloch Equations}

We follow \citet{Dicke1954} and approximate the Hamiltonian for a
sample of $N$ hydrogen atoms with each atom acting as a two-level
system, while taking into account the magnetic nature of the dipole-radiation
interaction applicable to this case,

\begin{equation}
\hat{H}=\hat{H}_{0}+\hat{H}_{\mathrm{rad}}+\hbar\sum_{j=1}^{N}\omega_{j}\left(\hat{R}_{j3}+\frac{\hat{1}}{2}\right)-\sum_{j=1}^{N}\hat{\mathbf{M}}_{j}\cdot\hat{\mathbf{\mathbf{B}}}\left(\mathbf{r}_{j}\right).\label{eq:Hamiltonian}
\end{equation}

\noindent In this Hamiltonian equation, $\hat{H}_{0}$ contains the
translational and \textcolor{black}{interatomic interaction energies}
of the atoms, $\hat{H}_{\mathrm{rad}}$ is the radiation field Hamiltonian
term, $\hbar\omega_{j}\left(\hat{R}_{j3}+\hat{1}/2\right)$ is the
internal energy of the $j^{\mathrm{th}}$ two-level atom ($\hat{1}$
is the unit operator), which has the eigenvalues $0$ and $\hbar\omega_{j}$,
and the last term stands for the interaction between the electromagnetic
field and the magnetic dipole of the $j^{\mathrm{th}}$ atom $\hat{\mathbf{M}}_{j}$.
Since this Hamiltonian is written under the magnetic dipole approximation,
it implies that\textcolor{red}{{} }\textcolor{black}{the magnetic field
$\hat{\mathbf{\mathbf{B}}}$ does not change }considerably over the
size of the atom, and is determined by its value at the position of
the center of mass of the atom, $\mathbf{r}_{j}$. Finally, the effects
of the hyperfine interaction between the proton and electron spins
within a single hydrogen atom and the Zeeman interaction due to an
external magnetic field would be included in the frequency $\omega_{j}$
of the atomic transition.

Following Dicke, we define the operators $\hat{R}_{x},\,\hat{R}_{y},\,\hat{R}_{3}$,
and $\hat{R}^{2}$ such that

\begin{eqnarray}
\hat{R}_{K}\left(\mathbf{r}\right) & = & \sum_{j=1}^{N}\hat{R}_{jK}\delta\left(\mathbf{r}-\mathbf{r}_{j}\right),\quad K=x,y,3\label{eq:RK}\\
\hat{R}^{2}\left(\mathbf{r}\right) & = & \hat{R}_{x}^{2}+\hat{R}_{y}^{2}+\hat{R}_{3}^{2}\label{eq:R2}\\
\left[\hat{R}^{2}\left(\mathbf{r}\right),\hat{R}_{K}\left(\mathbf{r^{\prime}}\right)\right] & = & 0\label{eq:=00005BR2,RK=00005D}\\
\left[\hat{R}_{a}\left(\mathbf{r}\right),\hat{R}_{b}\left(\mathbf{r}^{\prime}\right)\right] & = & i\varepsilon_{abc}\hat{R}_{c}\left(\mathbf{r}\right)\delta\left(\mathbf{r}-\mathbf{r}^{\prime}\right),\quad a,b,c=x,y,3,\label{eq:=00005BRa.Rb=00005D}
\end{eqnarray}

\noindent which are similar to the relations found in the spin or
general angular momentum formalisms. We can also define the raising
and lowering operators 
\begin{equation}
\hat{R}^{\pm}\left(\mathbf{r}\right)=\hat{R}_{x}(\mathbf{r})\pm i\hat{R}_{y}(\mathbf{r}),\label{eq:R+-}
\end{equation}
which further verify the following commutation relations

\begin{eqnarray}
\left[\hat{R}^{\pm}(\mathbf{r}),\hat{R}_{3}(\mathbf{r}')\right] & = & \mp\hat{R}^{\pm}(\mathbf{r})\delta(\mathbf{r}-\mathbf{r}')\label{eq:=00005BR+-,R3=00005D}\\
\left[\hat{R}^{+}(\mathbf{r}),\hat{R}^{-}(\mathbf{r}')\right] & = & 2\hat{R}_{3}(\mathbf{r})\delta(\mathbf{r}-\mathbf{r}').\label{eq:=00005BR+,R-=00005D}
\end{eqnarray}

\noindent It is clear from the form of the Hamiltonian and the commutation
relations between the operators that $\hat{R}^{2}$ and $\hat{R}_{3}$
commute with $\hat{H}$ and, therefore, share the eigenfunctions $|r,m_{r}\rangle$
introduced in Section \ref{sub:Dicke's-Small-sample-Model} to describe
the state of the system. 

The atomic hydrogen transitions at 21 cm are magnetic dipolar in nature
and bring into consideration the next leading term in our analysis,
i.e., the magnetic dipole interaction found on the right-hand-side
of Equation (\ref{eq:Hamiltonian})

\begin{equation}
\hat{V}_{\mathrm{MD}}=-\sum_{j=1}^{N}\hat{\mathbf{M}}_{j}\cdot\hat{\mathbf{B}}\left(\mathbf{r}_{j}\right),\label{eq:V_MD}
\end{equation}

\noindent which is at the center of our analysis. In general, the
magnetic dipole operator $\hat{\mathbf{M}}_{j}$ of the $j^{\mathrm{th}}$atom
can be written as \citep{Condon1935}

\begin{eqnarray}
\hat{\mathbf{M}}_{j} & = & \mu_{F}\hat{\mathbf{F}}_{j}\label{eq:M-F}\\
\mu_{F} & \simeq & g_{J}\left[\frac{F\left(F+1\right)+J\left(J+1\right)-I\left(I+1\right)}{2F\left(F+1\right)}\right],\label{eq:muF}
\end{eqnarray}

\noindent where $\hat{\mathbf{J}}$ is the sum of the electronic orbital
\textbf{$\hat{\mathbf{L}}$ }and spin $\hat{\mathbf{S}}$ angular
momenta (i.e., $\hat{\mathbf{J}}=\hat{\mathbf{L}}+\hat{\mathbf{S}}$),
and $\hat{\mathbf{F}}$ the sum of $\hat{\mathbf{J}}$ and the nuclear
spin $\hat{\mathbf{I}}$ (i.e., $\hat{\mathbf{F}}=\hat{\mathbf{J}}+\hat{\mathbf{I}}$).
For the hyperfine levels of the ground state of the hydrogen atom
we have $F=0\:\mathrm{and}\:1$, $J=S=1/2$, $I=1/2$, and $\mu_{F}\simeq g_{J}/2$,
whereas $g_{J}=g_{e}\mu_{\mathrm{B}}/\hbar$. In this equation, $g_{e}\simeq2$
and $\mu_{\mathrm{B}}$ is the Bohr magneton. The operator $\hat{\mathbf{F}}_{j}$
can also be written in terms of pseudo-spin operator $\hat{\mathbf{R}}_{j}$
as $\hat{\mathbf{F}}_{j}=\hbar\left(\hat{\mathbf{R}}_{j}+\hat{\mathbf{1}}/2\right)$,
allowing us to write 

\begin{equation}
\hat{\mathbf{M}}_{j}=\mu_{\mathrm{B}}\left(\hat{\mathbf{R}}_{j}+\frac{\hat{\mathbf{1}}}{2}\right).\label{eq:M(R)}
\end{equation}

For the $\left|F=0,m=0\right\rangle \longleftrightarrow\left|F=1,m=+1\right\rangle $
LCP transition we consider, the circular polarization state of radiation
can be defined using the corresponding unit vectors \citep{Grynberg2010}

\begin{eqnarray}
\mathbf{e}_{L} & = & -\frac{1}{\sqrt{2}}\left(\mathbf{e}_{x}+i\mathbf{e}_{y}\right)\label{eq:e+}\\
\mathbf{e}_{R} & = & \frac{1}{\sqrt{2}}\left(\mathbf{e}_{x}-i\mathbf{e}_{y}\right),\label{eq:e-}
\end{eqnarray}

\noindent which with ${\bf e}_{3}$ can be used to write the pseudo-spin
operator as

\begin{eqnarray}
\hat{{\bf R}}_{j} & = & \frac{1}{\sqrt{2}}\left(-\hat{R}_{j}^{-}{\bf e}_{L}+\hat{R}_{j}^{+}{\bf e}_{R}\right)+\hat{R}_{j3}{\bf e}_{3}.\label{eq:R-vector-definition}
\end{eqnarray}

In a H{\footnotesize I} gas, the LCP magnetic component of the radiation
propagating along the $\mathbf{k}$ direction interacts with the magnetic
dipole of a hydrogen atom resulting in a transition between the two
hyperfine levels. The corresponding magnetic field operator can be
expressed as

\begin{eqnarray}
\hat{{\bf B}}_{L}({\bf r},t) & = & \sum_{\mathbf{k}}\left[\hat{B}_{L\mathbf{k}}^{+}(\mathbf{r})e^{-i\omega_{k}t}\mathbf{e}_{L}+\hat{B}_{L\mathbf{k}}^{-}(\mathbf{r})e^{i\omega_{k}t}\mathbf{e}_{L}^{*}\right],\label{eq:B-operator}
\end{eqnarray}

\noindent where

\begin{eqnarray}
\hat{B}_{L\mathbf{k}}^{+}(\mathbf{r}) & = & \frac{1}{c}\sqrt{\frac{\hbar\omega_{k}}{2\epsilon_{0}V}}\hat{a}_{L\mathbf{k}}e^{i{\bf k}\cdot\mathbf{r}}\label{eq:B+quantized}\\
 & = & \left(\hat{B}_{L\mathbf{k}}^{-}\right)^{\dagger},\label{eq:B-quantized}
\end{eqnarray}
and $V$ is the arbitrary volume of quantization. In Equations (\ref{eq:B+quantized})
and (\ref{eq:B-quantized}), $a_{L\mathbf{k}}$ and $a_{L\mathbf{k}}^{\dagger}$
are, respectively, the LCP second quantization field annihilation
and creation operators, and obey the following commutation relation

\begin{equation}
\left[\hat{a}_{L\mathbf{k}},\hat{a}_{L\mathbf{k'}}^{\dagger}\right]=\hat{1}\delta_{\mathbf{kk}^{\prime}}.\label{eq:fieldcommutation}
\end{equation}

\noindent As a result, one can express the magnetic dipole interaction
term in Equation (\ref{eq:V_MD}) for transitions involving only LCP
photons as

\begin{eqnarray}
\hat{V}_{\mathrm{MD}} & = & -\mu_{\mathrm{B}}\sum\limits _{j=1}^{N}\hat{{\bf R}}_{j}\cdot\hat{\mathbf{B}}_{L}(\mathbf{r}_{j})\label{eq:V_MDj}\\
 & = & \frac{\mu_{\mathrm{B}}}{\sqrt{2}c}\sum_{j=1}^{N}\sum_{\mathbf{k}}\sqrt{\frac{\hbar\omega_{k}}{2\epsilon_{0}V}}\left(\hat{R}_{j}^{+}\hat{a}_{L\mathbf{k}}e^{i\mathbf{k}\cdot\mathbf{r}_{j}}+\hat{R}_{j}^{-}\hat{a}_{L\mathbf{k}}^{\dagger}e^{-i\mathbf{k\cdot}\mathbf{r}_{j}}\right),\label{eq:V_MDquantized}
\end{eqnarray}

\noindent using $\hat{\mathbf{R}}\cdot\mathbf{\mathbf{e}}_{L}=-\hat{R}^{+}/\sqrt{2}$
and $\hat{\mathbf{R}}\cdot\mathbf{e}_{L}^{*}=-\hat{R}^{-}/\sqrt{2}$.
It will also prove useful to write Equation (\ref{eq:V_MDquantized})
in the following form

\begin{equation}
\hat{V}_{\mathrm{MD}}=-\int_{V}\hat{\boldsymbol{\mathcal{M}}}(\mathbf{r})\cdot\hat{\mathbf{B}}_{L}(\mathbf{r})d^{3}r,\label{eq:V_MDmacro}
\end{equation}

\noindent which allows a definition of the transverse macroscopic
magnetization operator $\hat{\boldsymbol{\mathcal{M}}}(\mathbf{r})$
in terms of the raising and lowering density operators $\hat{R}^{+}$
and $\hat{R}^{-}$ as follows

\begin{eqnarray}
\hat{\boldsymbol{\mathcal{M}}}\left(\mathbf{r}\right) & {\color{red}{\normalcolor =}} & -\frac{\mu_{B}}{\sqrt{2}}\left[\hat{R}^{-}\left(\mathbf{r}\right)\mathbf{e}_{L}+\hat{R}^{+}\left(\mathbf{r}\right)\mathbf{e}_{L}^{*}\right]\label{eq:macroM}\\
{\color{red}} & \equiv & \hat{\boldsymbol{\mathcal{M}}}^{+}(\mathbf{r})+\hat{\boldsymbol{\mathcal{M}}}^{-}(\mathbf{r}).\label{eq:M+M-}
\end{eqnarray}

Neglecting inhomogeneous broadening effect (i.e., we omit $\hat{H}_{0}$
and set $\omega_{j}=\omega_{0}$) and inserting Equation (\ref{eq:V_MDquantized})
in Equation (\ref{eq:Hamiltonian}), the Hamiltonian of the H{\footnotesize I}-sample
system interacting with the 21 cm line via the ($F,m_{F}:0,0\longleftrightarrow1,+1$)
transition becomes 

\begin{equation}
\hat{H}=\hbar\omega_{0}\sum\limits _{j=1}^{N}\left(\hat{R}_{j3}+\frac{\hat{1}}{2}\right)+\hat{H}_{\mathrm{rad}}+\frac{\mu_{B}}{\sqrt{2}c}\sum_{\mathbf{k}}\sqrt{\frac{\hbar\omega_{\mathbf{k}}}{2\epsilon_{0}V}}\sum_{j=1}^{N}\left(\hat{R}_{j}^{+}\hat{a}_{L\mathbf{k}}e^{i\mathbf{k\cdot}\mathbf{r}_{j}}+\hat{R}_{j}^{-}\hat{a}_{L\mathbf{k}}^{\dagger}e^{-i\mathbf{k\cdot}\mathbf{r}_{j}}\right).\label{eq:Hamiltonian2}
\end{equation}

\noindent The radiation Hamiltonian term $\hat{H}_{\mathrm{rad}}$
can be expressed in terms of the second quantized operators $\hat{a}_{L\mathbf{k}}$
and $\hat{a}_{L\mathbf{k}}^{\dagger}$, and $\hat{a}_{R\mathbf{k}}$
and $\hat{a}_{R\mathbf{k}}^{\dagger}$ associated to left- and right-circular
polarized radiation states, respectively, with

\begin{equation}
\hat{H}_{\mathrm{rad}}=\hbar\omega\sum_{\mathbf{k}}\left[\left(\hat{a}_{L\mathbf{k}}^{\dagger}\hat{a}_{L\mathbf{k}}+\frac{\hat{1}}{2}\right)+\left(\hat{a}_{R\mathbf{k}}^{\dagger}\hat{a}_{R\mathbf{k}}+\frac{\hat{1}}{2}\right)\right].\label{eq:Hrad}
\end{equation}

As discussed in the literature \citep{Gross1982, Benedict1996} the
evolution of the atomic system can be calculated using the Heisenberg
equation of motion for the operator $\hat{X}$ in a system described
by the Hamiltonian $\hat{H}$ with 

\begin{equation}
\frac{d\hat{X}}{dt}=\frac{1}{i\hbar}[\hat{X},\hat{H}].\label{eq:Heisenberg}
\end{equation}

\noindent One can then readily find the following equations of motions
for $\hat{R}^{+}$, $\hat{R}^{-}$ and $\hat{R}_{3}$: 

\begin{eqnarray}
\frac{d\hat{R}^{+}}{dt} & = & i\omega_{0}\hat{R}^{+}-\frac{i\sqrt{2}\mu_{B}}{\hbar}\hat{R}_{3}\hat{B}_{L}^{-}\label{eq:R+t}\\
\frac{d\hat{R}^{-}}{dt} & = & -i\omega_{0}\hat{R}^{-}+\frac{i\sqrt{2}\mu_{B}}{\hbar}\hat{R}_{3}\hat{B}_{L}^{+}\label{eq:R-t}\\
\frac{d\hat{R}_{3}}{dt} & = & -\frac{i\mu_{B}}{\sqrt{2}\hbar}\left(\hat{R}^{+}\hat{B}_{L}^{+}-\hat{R}^{-}\hat{B}_{L}^{-}\right),\label{eq:R3t}
\end{eqnarray}

\noindent where, for simplicity, we now set $\hat{B}_{L}^{\pm}$ for
the value of the LCP-component of the magnetic field averaged over
the positions of the atoms. In a similar way, we can write the following
equations of motions for the raising and lowering magnetization operators
$\hat{\boldsymbol{\mathcal{M}}}^{+}(\mathbf{r})$ and $\hat{\boldsymbol{\mathcal{M}}}^{-}(\mathbf{r})$
using Equation (\ref{eq:macroM}) as

\begin{eqnarray}
\frac{d\hat{\boldsymbol{\mathcal{M}}}^{+}}{dt} & = & -i\omega_{0}\hat{\boldsymbol{\mathcal{M}}}^{+}-\frac{i\mu_{\mathrm{B}}^{2}}{\hbar}\left(\mathbf{e}_{L}\cdot\hat{\mathbf{B}}_{L}^{+}\right)\hat{R}_{3}\mathbf{e}_{L}\label{eq:M+t}\\
\frac{d\hat{\boldsymbol{\mathcal{M}}}^{-}}{dt} & = & i\omega_{0}\hat{\boldsymbol{\mathcal{M}}}^{-}+\frac{i\mu_{\mathrm{B}}^{2}}{\hbar}\left(\mathbf{e}_{L}^{*}\cdot\hat{\mathbf{B}}_{L}^{-}\right)\hat{R}_{3}\mathbf{e}_{L}^{*}.\label{eq:M-t}
\end{eqnarray}

\noindent It is also useful to define the operator $\hat{\mathcal{\mathbb{N}}}$

\begin{eqnarray}
\hat{\mathbb{N}} & = & 2\hat{R}_{3}\label{eq:N-def}\\
 & = & 2\sum_{j=1}^{N}\hat{R}_{3j}\delta(\mathbf{r}-\mathbf{r}_{j}),\label{eq:N-def-micro}
\end{eqnarray}

\noindent which can be interpreted as a population inversion density
operator considering that $\hat{R}_{3j}$ has eigenvalues of $\pm1/2$
and the eigenvalue of $\hat{R}_{3}=\sum\hat{R}_{3j}$ is equal to
half of the population difference between the excited level $F=1,\,m_{F}=1$
and the ground level $F=0,\,m_{F}=0$ at time $t$. Using (\ref{eq:N-def-micro})
one can show that Equations (\ref{eq:R3t})-(\ref{eq:M-t}) can be
rewritten as

\begin{eqnarray}
\frac{d\hat{\mathbb{N}}}{dt} & = & \frac{2i}{\hbar}\left(\hat{\boldsymbol{\mathcal{M}}}^{+}\cdot\hat{\mathbf{B}}_{L}^{+}-\hat{\boldsymbol{\mathcal{M}}}^{-}\cdot\hat{\mathbf{B}}_{L}^{-}\right)\label{eq:N-t}\\
\frac{d\hat{\boldsymbol{\mathcal{M}}}^{+}}{dt} & = & -i\omega_{0}\hat{\boldsymbol{\mathcal{M}}}^{+}-i\gamma\left(\mathbf{e}_{L}\cdot\hat{\mathbf{B}}_{L}^{+}\right)\hat{\mathbb{N}}\mathbf{\mathbf{e}}_{L}\label{eq:M+t-final}\\
\frac{d\hat{\boldsymbol{\mathcal{M}}}^{-}}{dt} & = & i\omega_{0}\hat{\boldsymbol{\mathcal{M}}}^{-}+i\gamma\left(\mathbf{e}_{L}^{*}\cdot\hat{\mathbf{B}}_{L}^{-}\right)\hat{\mathbb{N}}\mathbf{\mathbf{e}}_{L}^{*},\label{eq:M-t-final}
\end{eqnarray}

\noindent where $\gamma=\mu_{\mathrm{B}}^{2}/2\hbar$. 

\noindent Furthermore, in the Heisenberg representation one can derive
the following equation 

\begin{equation}
-\nabla^{2}\hat{\mathbf{B}}_{L}^{\pm}+\frac{1}{c^{2}}\frac{\partial^{2}\hat{\mathbf{B}}_{L}^{\pm}}{\partial t^{2}}=-\mu_{0}\nabla^{2}\hat{\boldsymbol{\mathcal{M}}}^{\pm}\label{eq:B-t-vec}
\end{equation}
for the evolution of the magnetic component of the radiation field
when defining 

\begin{eqnarray}
\hat{B}_{L}^{\pm}\left(\mathbf{r},t\right) & = & \hat{B}_{0}^{\pm}\left(\mathbf{r},t\right)e^{\pm i\left(kz-\omega t\right)}\label{eq:B-wave}\\
\hat{\mathcal{M}}^{\pm}\left(\mathbf{r},t\right) & = & \hat{\mathcal{M}}_{0}^{\pm}\left(\mathbf{r},t\right)e^{\pm i\left(kz-\omega t\right)},\label{eq:M-wave}
\end{eqnarray}

\noindent using the slowly varying envelope approximation (SVEA),
where $\hat{B}_{0}^{\pm}$ and $\hat{\mathcal{M}}_{0}^{\pm}$ are
slow varying envelope operators multiplied by fast oscillating exponential
terms propagating in the positive $z$ direction. Within the context
of the SVEA we assume that the $\hat{B}_{0}^{\pm}$ and $\hat{\mathcal{M}}_{0}^{\pm}$
significantly change over time-scales much longer than $1/\omega$
and length-scales much larger than $1/k$ \citep{Gross1982}. Upon
applying the SVEA, Equation (\ref{eq:B-t-vec}) is simplified to

\noindent 
\begin{eqnarray}
\left(\frac{\partial}{\partial z}+\frac{1}{c}\frac{\partial}{\partial t}\right)\hat{B}_{0}^{\pm} & \simeq & \pm\frac{i\mu_{0}\omega}{2c}\hat{\mathcal{M}}_{0}^{\pm}.\label{eq:final-B-z-t}
\end{eqnarray}
In the derivation of Equation (\ref{eq:final-B-z-t}), we neglected
any transverse effects on the radiation field and magnetization (i.e.,
$\partial B_{0}^{\pm}/\partial x\approx\partial B_{0}^{\pm}/\partial y\approx0$
and $\partial\hat{\mathcal{M}}_{0}^{\pm}/\partial x\approx\partial\hat{\mathcal{M}}_{0}^{\pm}/\partial y\approx0$),
which should be included in numerical calculations for a true three-dimensional
sample.

Using Equations (\ref{eq:B-wave}) and (\ref{eq:M-wave}) we can rewrite
Equations (\ref{eq:N-t}), (\ref{eq:M+t-final}), and (\ref{eq:M-t-final})
at resonance, i.e., when $\omega=\omega_{0}$, in the reduced form
of 

\begin{eqnarray}
\frac{d\hat{\mathcal{M}}_{0}^{\pm}}{dt} & = & \mp i\gamma\hat{B}_{0}^{\pm}\hat{\mathbb{N}}\label{eq:final-ideal-M-t}\\
\frac{d\hat{\mathbb{N}}}{dt} & = & \frac{2i}{\hbar}\left(\hat{\mathcal{M}}_{0}^{-}\hat{B}_{0}^{+}-\hat{\mathcal{M}}_{0}^{+}\hat{B}_{0}^{-}\right).\label{eq:final-ideal-N-t}
\end{eqnarray}

\noindent Equations (\ref{eq:final-B-z-t}), (\ref{eq:final-ideal-M-t})
and (\ref{eq:final-ideal-N-t}) are known as the Maxwell-Bloch equations,
and can be solved simultaneously to determine the time evolution of
the radiation field, magnetization, and excitation state for an ideal
sample.

\subsection{Dephasing Effects and Pumping\label{sub:Dephasing-Effects}}

The previous derivations for the ideal case must be augmented appropriately
when dealing with more realistic conditions for the ISM, where dephasing
and relaxation effects cannot be neglected and continuous pumping
of the atomic system can take place. One can phenomenologically add
the corresponding terms to the atomic equations as follows \citep{Mandel2010}
\begin{eqnarray}
\frac{d\hat{\mathcal{M}}_{0}^{+}}{dt} & = & -i\gamma\hat{B}_{0}^{+}\hat{\mathbb{N}}-\frac{1}{T_{2}}\hat{\mathcal{M}}_{0}^{+}+\Lambda_{\mathrm{M}}\label{eq:finalM+}\\
\frac{d\hat{\mathbb{N}}}{dt} & = & \frac{2i}{\hbar}\left(\hat{\mathcal{M}}_{0}^{-}\hat{B}_{0}^{+}-\hat{\mathcal{M}}_{0}^{+}\hat{B}_{0}^{-}\right)-\frac{1}{T_{1}}\left(\hat{\mathbb{N}}-\mathbb{N_{\mathrm{eq}}}\right),\label{eq:finalN}
\end{eqnarray}

\noindent where $T_{1}$ and $T_{2}$ are the characteristic time-scales
for, respectively, population decay and de-magnetization, $\mathbb{N}_{\mathrm{\mathrm{eq}}}$
is the ``equilibrium'' value for $\hat{\mathbb{N}}$ obtained in
the absence of interaction with the coherent field $\hat{B}_{L}$,
while $\Lambda_{\mathrm{M}}$ represents any source term of magnetization. 

The one-dimensional magnetic field Equation (\ref{eq:final-B-z-t})
can also be adapted to the more realistic conditions by adding a correction
term to account for the loss of radiation due to transverse effects
and diffraction, which depend on the shape and symmetry of the sample.
These are characterized by the Fresnel number 
\begin{equation}
F_{\mathrm{n}}=\frac{A}{\lambda L},\label{eq:Fresnel}
\end{equation}
where $A$ and $L$, respectively, stand for the cross-section and
length of the sample. For samples of cylindrical symmetry with Fresnel
number smaller than one, transverse effects of the field are negligible,
whereas the diffraction of radiation along the propagation axis can
play an important role. \citet{Gross1982} have shown that a damping
term $\textit{B}_{0}^{+}/L_{\mathrm{diff}}$ can be included in the
field equation to take into account diffraction effects in samples
with $F_{\mathrm{n}}\ll1$. For such a sample Equation (\ref{eq:final-B-z-t})
can be approximately augmented to

\begin{equation}
\left(\frac{\partial}{\partial z}+\frac{1}{c}\frac{\partial}{\partial t}\right)\hat{B}_{0}^{+}\left(z,\tau\right)+\frac{1}{L_{\mathrm{diff}}}B_{0}^{+}\left(z,\tau\right)\simeq\frac{i\mu_{0}\omega}{2c}\hat{\mathcal{M}}_{0}^{+},\label{eq:finalfield}
\end{equation}
where $L_{\mathrm{diff}}\simeq F_{\mathrm{n}}L/0.35$ \citep{Gross1982}.

The atom-field Equations (\ref{eq:finalM+}), (\ref{eq:finalN}),
and (\ref{eq:finalfield}) also form a Maxwell-Bloch system of equations,
and provide a more complete and realistic picture for the evolution
of the system. This set of equations can be numerically solved for
a given set of parameters $T_{1},\:T_{2},\;\mathbb{N}_{\mathrm{eq}}$
and $\Lambda_{\mathrm{M}}$.

\section{The Sine-Gordon Solution\label{sec:Non-linear-regime}}

The set of Equations (\ref{eq:finalM+}), (\ref{eq:finalN}) and (\ref{eq:finalfield})
can only be solved analytically for a few special cases \citep{Mandel2010}.
We first consider the ideal condition, where the dephasing/relaxation,
diffraction, and pumping terms are neglected (i.e., $T_{1}=T_{2}=\infty$,
$L_{\mathrm{diff}}=\infty$, and $\Lambda_{\mathrm{M}}=0$). Effecting
a change of variable from $t$ to the retarded time $\tau=t-z/c$
yields

\begin{eqnarray}
\left(\frac{\partial}{\partial z}+\frac{1}{c}\frac{\partial}{\partial t}\right) & = & \frac{\partial}{\partial z}\label{eq:d/dz}\\
\frac{\partial}{\partial t} & = & \frac{\partial}{\partial\tau},\label{eq:d/dtau}
\end{eqnarray}

\noindent which can be used to simplify the set of Maxwell-Bloch equations
(\ref{eq:final-B-z-t})-(\ref{eq:final-ideal-N-t}) to

\begin{eqnarray}
\frac{\partial\hat{\mathcal{M}}_{0}^{\pm}}{\partial\tau} & = & \mp i\gamma\hat{B}_{0}^{\pm}\hat{\mathbb{N}}\label{eq:ideal-M-tau}\\
\frac{\partial\hat{\mathbb{N}}}{\partial\tau} & = & \frac{2i}{\hbar}\left(\hat{\mathcal{M}}_{0}^{-}\hat{B}_{0}^{+}-\hat{\mathcal{M}}_{0}^{+}\hat{B}_{0}^{-}\right)\label{eq:ideal-N-tau}\\
\frac{\partial\hat{B}_{0}^{\pm}\left(z,\tau\right)}{\partial z} & \simeq & \pm\frac{i\mu_{0}\omega}{2c}\hat{\mathcal{M}}_{0}^{\pm}.\label{eq:ideal-field-tau-z}
\end{eqnarray}

The form of Equations (\ref{eq:ideal-M-tau}) and (\ref{eq:ideal-N-tau})
implies that $\left|\hat{\boldsymbol{\mathcal{M}}}^{+}\right|^{2}+\left|\hat{\boldsymbol{\mathcal{M}}}^{-}\right|^{2}+\left(\mu_{\mathrm{B}}^{2}/4\right)\left|\hat{\mathbb{N}}\right|^{2}$
is a conserved quantity and allows us to redefine $\hat{\mathcal{M}}_{0}^{\pm}$
and $\mathbb{N}$ as

\noindent 
\begin{eqnarray}
\hat{\mathcal{M}}_{0}^{+} & = & \frac{\mu_{\mathrm{B}}N}{2\sqrt{2}V}\sin\left(\theta\right)\label{eq:M+-spheric}\\
\hat{\mathbb{N}} & = & \frac{N}{V}\cos\left(\theta\right),\label{eq:Nspheric}
\end{eqnarray}
where $N$ is the number of inverted atoms in the sample at $\tau=0$
and $\theta$ is the so-called Bloch angle.

Taking these solutions into account, at resonance Equations (\ref{eq:ideal-M-tau})
and (\ref{eq:ideal-field-tau-z}) are transformed to 

\begin{eqnarray}
\hat{B}_{0}^{+} & = & \frac{i\mu_{\mathrm{B}}}{2\sqrt{2}\gamma}\frac{\partial\theta}{\partial\tau}\label{eq:B-theta-ideal}\\
\frac{\partial\hat{B}_{0}^{+}}{\partial z} & = & \frac{i\mu_{0}\omega\mu_{\mathrm{B}}N}{4\sqrt{2}cV}\sin\left(\theta\right)\label{eq:B-z-sin-theta}
\end{eqnarray}

\noindent in the retarded-time frame. Taking the spatial derivative
of Equation (\ref{eq:B-theta-ideal}) we can write

\begin{equation}
\frac{\partial\hat{B}_{0}^{+}}{\partial z}=\frac{i\mu_{\mathrm{B}}}{2\sqrt{2}\gamma}\frac{\partial^{2}\theta}{\partial z\partial\tau},\label{eq:derivativeB0}
\end{equation}

\noindent which when compared to Equation (\ref{eq:B-z-sin-theta})
yields the following non-linear equation

\begin{equation}
\frac{\partial^{2}\theta}{\partial z\partial\tau}=\frac{\mu_{0}\mu_{\mathrm{B}}^{2}\omega N}{4\hbar cV}\sin\left(\theta\right)\label{eq:theta}
\end{equation}

\noindent upon using $\gamma=\mu_{\mathrm{B}}^{2}/2\hbar$. This equation
is further transformed with the introduction of a new dimensionless
variable \citep{Gross1982} 
\begin{equation}
q=2\sqrt{\frac{z\tau}{LT_{\mathrm{R}}}},\label{eq:q}
\end{equation}

\noindent to 

\begin{equation}
\frac{d^{2}\theta}{dq^{2}}+\frac{1}{q}\frac{d\theta}{dq}=\sin\left(\theta\right),\label{eq:sine-gordon}
\end{equation}

\noindent with $T_{\mathrm{R}}$ the characteristic time for superradiance
given by Equation (\ref{eq:TR-largesample}). Equation (\ref{eq:sine-gordon})
is the so-called Sine-Gordon equation \citep{Gross1982}. This equation
can be numerically solved and the corresponding solution for $\theta$
substituted back into Equation (\ref{eq:B-theta-ideal}) to determine
the field amplitude $\hat{B}_{0}^{+}$ emerging out of the sample
(at $z=L$) as a function of the retarded time $\tau$. Knowing $\hat{B}_{0}^{+}\left(L,\tau\right)$,
the output radiation intensity\textbf{ $I$ }is given by

\begin{equation}
I=\frac{c}{2\mu_{0}}\left|\hat{B}_{0}^{+}\right|^{2}.\label{eq:intensityequation}
\end{equation}

A more realistic case where dephasing/relaxation is included with
a single time-scale (i.e., $T^{\prime}=T_{1}=T_{2}\neq\infty$, $L_{\mathrm{diff}}=\infty$,
and $\Lambda_{\mathrm{M}}=\mathbb{N_{\mathrm{eq}}}=0$) can be dealt
with in a similar manner. We then have the corresponding definitions
for Equations (\ref{eq:M+-spheric}) and (\ref{eq:Nspheric})

\begin{eqnarray}
\hat{\mathcal{M}}_{0}^{+} & = & \frac{\mu_{\mathrm{B}}N}{2\sqrt{2}V}\sin\left(\theta\right)e^{-\tau/T^{\prime}}\label{eq:Rcursive_M-dephase}\\
\hat{\mathbb{N}} & = & \frac{N}{V}\cos\left(\theta\right)e^{-\tau/T^{\prime}},\label{eq:N_dephase}
\end{eqnarray}

\noindent which also lead to Equation (\ref{eq:B-theta-ideal}) for
$\hat{B}_{0}^{+}$. Performing a spatial derivative on Equation (\ref{eq:B-theta-ideal})
yields 

\begin{equation}
\frac{\partial^{2}\theta}{\partial z\partial\tau}=\frac{\mu_{0}\mu_{\mathrm{B}}^{2}\omega N}{4\hbar cV}\sin\left(\theta\right)e^{-\tau/T^{\prime}}.\label{eq:theta2}
\end{equation}

\noindent A comparison of Equations (\ref{eq:theta2}) with (\ref{eq:theta})
shows that the presence of dephasing implies a source term containing
a decaying exponential. This exponential factor can be removed from
this equation through the following change of variable 

\begin{equation}
\tau\longrightarrow\tau^{\prime}=T^{\prime}\left(1-e^{-\tau/T^{\prime}}\right),\label{eq:tau_prime}
\end{equation}

\noindent which allows us to transform Equation (\ref{eq:theta2})
to the Sine-Gordon equation (i.e., Equation (\ref{eq:sine-gordon}))
by redefining the dimensionless parameter $q$ with the following

\begin{equation}
q=2\sqrt{\frac{z\tau^{\prime}}{LT_{\mathrm{R}}}}.\label{eq:newq}
\end{equation}

\end{document}